\documentclass[aps,prb,twocolumn,superscriptaddress,showpacs]{revtex4}

\usepackage{amsmath}
\usepackage{graphicx}
\usepackage[usenames,dvipsnames]{color}

\begin{document}

\title{ Benchmarking Hydrogen-Helium Mixtures with QMC:  Energetics, Pressures, and Forces}

\author{Raymond C. Clay III}
\email[]{rcclay2@illinois.edu}
\affiliation{Department of Physics, University of Illinois at Urbana-Champaign, Urbana, Illinois 61801, USA}

\author{Markus Holzmann}
\affiliation{LPMMC, UMR 5493 of CNRS, Universit{\'e} Grenoble Alpes, F-38100 Grenoble, France}
\author{David M. Ceperley}
\affiliation{Department of Physics, University of Illinois at Urbana-Champaign, Urbana, Illinois 61801, USA} 

\author{Miguel A. Morales}
\affiliation{Lawrence Livermore National Laboratory, Livermore, California 94550, USA}

\date{\today}

\begin{abstract}
An accurate understanding of the phase diagram of dense hydrogen and helium mixtures is a crucial component in the construction of accurate models of Jupiter, Saturn, and Jovian extrasolar planets.  Though DFT based first principles methods have the potential to provide the accuracy and computational efficiency required for this task, recent benchmarking in hydrogen has shown that achieving this accuracy requires a judicious choice of functional, and a quantification of the errors introduced.  In this work, we present a quantum Monte Carlo based benchmarking study of a wide range of density functionals for use in hydrogen-helium mixtures at thermodynamic conditions relevant for Jovian planets.  Not only do we continue our program of benchmarking energetics and pressures, but we deploy QMC based force estimators and use them to gain insights into how well the local liquid structure is captured by different density functionals.  We find that TPSS, BLYP and vdW-DF are the most accurate functionals by most metrics, and that the enthalpy, energy, and pressure errors are very well behaved as a function of helium concentration.  Beyond this, we highlight and analyze the major error trends and relative differences exhibited by the major classes of functionals, and estimate the magnitudes of these effects when possible.  
\end{abstract}



\maketitle

\section{Introduction}

Since the first exoplanet was discovered in 1988, the advancements in observational techniques and the launch of the Kepler telescope have revolutionized our understanding of other solar systems.  To date, there are almost 2000 catalogued extrasolar planets \cite{Schneider2011}, a large fraction of which are Jovian type gas giants and brown dwarves whose compositions are more than 90\% hydrogen and helium--similar to Jupiter and Saturn.  Understanding the evolution of these extrasolar planets and their parent star systems would thus be greatly aided by accurate models.  Current planetary models for Jovian planets have the advantage that by knowing a scant few observables, like the mass, radius, luminosity, and element compositions, one can determine the interior structure and time evolution of the planet.  However, in order to construct accurate models, one needs an accurate equation of state for hydrogen-helium mixtures at all temperatures, pressures, and species concentrations relevant in planetary interiors.  

In the absence of experimental data at these extreme conditions, perturbative methods \cite{Stevenson1975}, chemical models \cite{Saumon1995} and  \textit{ab initio} methods \cite{Klepeis1991, Pfaffenzeller1995, Lorenzen2009, Lorenzen2011, Morales2009, Morales2013b} have done an excellent job in identifying phases that could be relevant in planetary models.  For instance, the metallization of dense hydrogen liquids is universally believed to occur at temperatures and pressures appropriate to the core regions of Jovian planets, and should be the source of the planetary dynamo responsible for their large magnetic fields.  More subtly is the possibility of immiscibility of helium in dense hydrogen liquids, whereby homogeneous H+He mixtures driven into the upper atmosphere would condense into helium rich droplets, which would rain down to the deeper atmosphere\cite{Stevenson1977,Stevenson1977a}.  This process could provide an additional energy dissipation mechanism as well as fundamentally alter the mass distribution of helium and heavier elements in the upper atmosphere, and so should be treated accurately in planetary models.  

Unfortunately, there is enough quantitative and qualitative disagreement among chemical model and \textit{ab initio} based calculations regarding the locations and nature of the metallization and immiscibility transitions that planetary scientists routinely treat the equation of state as a free parameter in their models--to be varied to match the experimental data\cite{Fortney2003,Nettelmann2015}.  Recent models constructed in this manner show an excellent ability to reproduce the observed atmospheric depletion of He and excess luminosity in Saturn, the gravitational moments, and estimated ages of both Jupiter and Saturn \cite{Fortney2003,Nettelmann2015}.  The unfortunate cost of this approach is that it leaves large uncertainties in other areas of the model--for example the core mass and composition, the distribution of heavier elements throughout the planet, etc.  In principle, these uncertainties could be greatly reduced by the additional constraints imposed by an accurate equation of state.  

To move forward with the construction of an accurate \textit{ab initio} equation of state with well established error estimates, one needs to understand and accommodate for two frequently made approximations.  The first is ideal mixing. The validity of the ideal mixing approximation for the entropy, which has been used in chemical models and ab initio calculations since the 1970's, has only recently been investigated in the context of ab initio simulations using thermodynamic integration (TI) techniques \cite{Morales2013b, Morales2009}.  There is a current quantitative discrepancy of approximately 1000K in the demixing transition at high pressures, and significant qualitative disagreement at lower pressures between the works of Morales et al. and Lorenzen et al \cite{Lorenzen2009, Lorenzen2011}.  Fortunately, with current computational resources, this source of error can be effectively eliminated through the use of TI.  

The second, and least understood source of error is in the treatment of electronic correlation effects, typically through the use of an approximate exchange correlation functional within density functional theory (DFT).  In pure hydrogen, one can at least quantify the impact of density functional errors on the phase diagram, both because of explicit benchmarking studies, and because the phase diagram has been computed by many different groups and methods, ranging from slews of different functionals to highly accurate QMC methods.  Various studies have found that the pure hydrogen phase diagram is extremely sensitive to the choice of exchange correlation functional, primarily because of the presence of multiple molecular disassociation and metallization phase transitions and crossovers\cite{Morales2013, Morales2013a, Clay2014, McMinis2015}.  In contrast, little is known of the impact that density functionals errors have on the demixing temperature in dense H+He mixtures.  Since the mid 90's, PBE has become more favored than LDA, but the methodological and quantitative differences between various studies precludes a direct comparison.  That dense H+He mixtures can demonstrate similar metallization and quasi-molecular transitions should inspire caution.

Alternative methods for solving the electronic structure exist.  In particular, projector Quantum Monte Carlo (p-QMC) is a highly accurate, variational, many-body method, well suited to treating electronic correlation in low Z materials.  Hydrogen bonding, van der Waals forces, and other types of difficult electronic effects are automatically taken into account.  The drawback of using QMC in computing entire phase diagrams is that it is about two orders of magnitude more expensive than DFT calculations, which makes its widespread use in molecular dynamics and ionic Monte Carlo applications comparatively unattractive with today's computational resources.  Moreover, it is significantly more difficult to compute more complex properties relevant to planetary physics applications within the QMC framework, like electrical conductivity and viscosity.  

In this work, we instead propose to use projector Quantum Monte Carlo (QMC) methods to benchmark a range of density functionals in thermodynamic regimes relevant for helium sedimentation in Jovian planets.  Our main objective is to identify and understand qualitative error trends and relative differences between various classes of density functionals exhibit when used to estimate thermodynamic quantities in hydrogen-helium mixtures.   Though we in many cases strive for quantitative accuracy in addition, this should not distract from our main objective.  To achieve this, we benchmark the errors occurring in the energetics, pressures, and forces for each functional, and note how they change as a function of both density and helium concentration. 

This article is organized as follows.  In section \ref{sec:comp_details}, we first discuss the computational details specific to our current work.  Then in section \ref{sec:results} we present the benchmarking results for global and local energetics, pressures, enthalpies, and forces.  In section \ref{sec:discussion}, we explain the error trends we observe in terms of the underlying exchange functional, after which we conclude.  


\section{Method} 
\label{sec:comp_details}
In this study, we employ the same general methodology we used previously in our work on pure hydrogen \cite{Clay2014}.  Thus, we will in this section focus strictly on simulation details and extensions to the method, leaving the high level justification and detailed explanation of the approach to our previous paper. 
\subsection{Test Sets }
The relevant thermodynamic variables for describing the H+He phase diagram are the density $\rho$, the temperature $T$, and the helium species fraction $x_{He}$, which is defined as:
\begin{equation}
x_{He}=\frac{N_{He}}{N_H+N_{He}}
\end{equation}

In the study of dense hydrogen and helium, it is often customary to list densities in terms of $r_s$ instead of $\rho$, which are related by $\Omega/N_e = \frac{4}{3}\pi r_s^3$.  $N_e$ is the total number of electrons in our system and $\Omega$ is the volume.  

As our goal is to establish the accuracy of density functionals around the demixing transition in Saturn and Jupiter, all test sets are chosen along the T=7000K isotherm.  We considered three different densities: $r_s=1.10, 1.25, 1.34$. At each density, in addition to the pure hydrogen and pure helium cases, we considered helium concentrations between $0-20.7\%$, since these are the most relevant compositions for planetary interiors. 

Samples at these thermodynamic conditions consisted of charge-neutral cubic cells of 64 electrons with differing numbers of H and He ions to ensure charge neutrality.  At each density and helium concentration, twenty statistically independent samples were generated from ab-initio quantum molecular dynamics simulations in the NVT ensemble using classical nuclei.  The MD simulations were performed with the VASP \cite{Kresse1993, Kresse1994,Kresse1996,Blochl1994} simulation package using the PBE \cite{Perdew1996} exchange correlation functional. 

\subsection{Density Functional Comparison}\label{sec:DFTcomparison}
For all configurations, we calculated the total energy, stresses, and forces using the following functionals in Quantum Espresso\cite{Giannozzi2009}: LDA\cite{Perdew1981}; (\textit{GGA}) PBE, revPBE\cite{Zhang1998}, PBEsol\cite{Perdew2008}, BLYP\cite{Lee1988}, Wu-Cohen\cite{Wu2006}, (metaGGA) M06L\cite{Zhao2006}, TPSS \cite{Tao2003}; (\textit{non-local dispersion corrected}) vdW-DF \cite{Dion2004}, vdW-DF2 \cite{Lee2010},  vdW-DF-C09, vdW-DF2-C09 \cite{Cooper2010}, vdW-DF-CX\cite{Berland2014}, vdW-optB86B \cite{Klimes2011}, and vdW-optB88 \cite{Klimes2010}.

For all above functionals, we used a plane wave cutoff of 800 Ry and a $7x7x7$ Monkhorst-Pack grid with an offset.  We used hard Troullier-Martin pseudopotentials\cite{Troullier1991} with no core electrons for both H and He that were generated with Opium \cite{OPIUM} using the PBE functional.  To ensure no pseudopotential overlap in our test-set, we chose real space cutoffs of $r_c=0.37a_0$ and $r_c=0.5a_0$ for the H and He pseudopotentials respectively.  

We also tested the exact-exchange HSE functional, however due to computational and memory limitations, we took the following cost saving measures.  First, at every density and helium concentration we considered, we performed only 10 HSE calculations.  Secondly, we reduced the Monkhorst-Pack grid to 5x5x5 (the same grid used to evaluate the Fock operator), and ran the calculations with VASP because of its compact PAW formalism.  A planewave cutoff of 1500eV and 96 bands were used for all calculations.  We found that this gives the desired accuracy for energy and pressure differences within configurations at the same density and helium concentration, however comparisons between different densities and helium concentrations might be slightly underconverged.  Since we had a limited choice of pseudopotentials, we were unable to perform calculations at $r_s=1.10$ and guarantee the desired level of accuracy for VASP calculations.  

\subsection{Quantum Monte Carlo calculations}

All quantum Monte Carlo calculations were done using the QMCPACK \cite{Esler2012a, Kim2012} simulation package.   The trial wavefunction is taken to be of the single Slater-Jastrow form.  Singe particle orbitals were obtained from Quantum Espresso \cite{Giannozzi2009} using the PBE functional, the same Troullier-Martin pseudopotentials described previously, and a planewave cutoff of 200Ry.  We used pseudopotentials only in the orbital generation step to eliminate the electron-ion cusp, which we preferred to handle within the electron-ion Jastrow terms.  All QMC calculations were ``all-electron": we used the bare coulomb interaction between electrons and electrons, and between electrons and nuclei.  

For the Jastrow factor, we used short-ranged one-body and two-body functions of b-spline form.  For H and He, the one-body terms were spin-independent.  The one body term for each species was a sum of two functions. The first was a ``core" jastrow, which had a real space cutoff of $r_c=1.0 a_0$,  8 knots, and had the suitable electron-ion cusp condition imposed.  The second had a cutoff of $L/2$ with 8 knots and no cusp-condition imposed.  For the two-body functions, we separately included same-spin and opposite-spin e-e terms, each with a cutoff of $r_c=L/2$ and correct cusp conditions imposed.  

Our wavefunctions were optimized with the linear method\cite{Umrigar2007}.  After obtaining a good initial guess for the jastrow parameters from a single $r_s=1.10$ configuration with 4 He atoms, all variational parameters were simultaneously optimized using an initial variance minimization step, followed by 10 energy minimization steps.  Convergence of the minimization procedure was checked.

Energies, pressures, and the structure factor were calculated using Reptation Monte Carlo (RMC) \cite{Baroni1999, Pierleoni2005}.  Our target statistical error bars for the energies and pressures were 0.008 mHa/electron and 0.3GPa respectively.  For all but the pure helium configurations, we used a time-step of $\tau=0.0075 Ha^{-1}$ and projection time of $\beta=4.5 Ha^{-1}$.  These choices were found to yield time-step and mixed-estimator errors for the potential energy that were comparable to the desired error bars.  For the pure helium configurations, we fixed the projection time at $\beta=4.5 Ha^{-1}$ and ran with time steps of $\tau=0.0075 Ha^{-1}$ and $\tau=0.00375 Ha^{-1}$.  All accumulated quantities were then linearly extrapolated to zero time step.

Forces were computed using the Chiesa, Ceperley, Zhang estimator \cite{Chiesa2005} adapted to periodic boundary conditions, which we detail in the supplemental information.  We used a real-space cutoff of $\mathcal{R}=1.0 a_0$ and a smoothing polynomial of degree $M=3$.  Based on several statistical tests detailed in the supplemental information, we found that this choice of parameters yielded systematic errors that were less than the error bar on the hydrogen force components, which was approximately 2mHa/bohr. This resolution was sufficient to clearly distinguish different functionals.  We used diffusion Monte Carlo with a time-step of $\tau=0.01 Ha^-1$ and a population size of 512 walkers, which we found converged the local energy (but not the potential) to within error bars.  To correct for the mixed-estimator problem we used extrapolated force estimates.  All systematic errors are expected to be less than the statistical error bar.  


We applied the following finite size corrections.  To reduce shell effects, we used canonical twist-averaged boundary conditions (TABC) on a 4x4x4 Monkhorst-Pack grid \cite{Lin2001}.  For the potential energy correction, we used the leading order Chiesa correction based on pure estimates of $S_{ee}(k)$ \cite{Chiesa2006}.  For the kinetic energy correction, we detail this in the supplemental information.  Since most configurations are in the metallic state, there is also a kinetic energy error arising from the fact that we are attempting to reproduce a fermi-surface with only 64 electrons.  To correct for this, we used the PBE functional and estimated the energy error between a twist-averaged unit cell and a 7x7x7 MP grid.  This correction scheme was tested against several supercell calculations at $r_s=1.10$ and $r_s=1.34$ across all helium concentrations.  We expect the absolute energy errors  (across all densities and helium concentrations) to be less than 0.5mHa/electron, and the pressure errors to be approximately 1GPa.  Details are in the supplemental information.  

Regarding QMC force errors, a few caveats are worth mentioning.  No finite-size correction scheme beyond twist-averaging was applied to the forces.  Though this should eliminate the bulk of the finite-size error, there is still a residual finite size error coming from the fact that $\rho(\mathbf{r})$ is not quite converged to the thermodynamic limit.  By comparing the forces from a twist-averaged KZK calculation in the unit cell with those of a close to converged 7x7x7 MP grid, we estimate that the standard deviation of the residual finite-size error in the force estimates is approximately 2mHa/bohr. 
\subsection{Error Analysis}
\subsubsection{Scalar Quantities}
As in our previous paper, we consider a test set $S$ with $M$ configurations, $\lbrace \mathbf{R}_0 \ldots \mathbf{R}_M \rbrace$.  For each configuration $\mathbf{R}_i$, we define a density functional error $\delta \mathcal{A}(\mathbf{R}_i)=\mathcal{A}^{DF}(\mathbf{R}_i)-\mathcal{A}^{QMC}(\mathbf{R}_i)$, where $\mathcal{A}$ is some observable (e.g. total energy, pressure), ``DF" is the density functional, ``QMC" is the QMC reference value.  

In addition to defining average errors over a test set S, which we denote $\langle \delta \mathcal{A}^{DF} \rangle_S$, we define a general class of shifted mean absolute errors as:
\begin{equation}\label{error_measure}
\langle |\widetilde{\delta \mathcal{A}}| \rangle_S = \frac{1}{M}\sum_{\mathbf{R}_i \in S} |\delta \mathcal{A}^{DF}(\mathbf{R}_i) - c^{DF} |
\end{equation}
Here, $c^{DF}$ is an density functional dependent offset.  The standard ``mean absolute error" corresponds to the choice of $c^{DF}=0$, which we will label as $\langle |\delta \mathcal{A}| \rangle_S$.  However, we will in this paper find cause to use other choices of $c^{DF}$ for different observables, specifically for measures of ``global" and ``local" energetics, which we will take care to explain as they arise. 

\subsubsection{Forces}\label{sec:force_defs}
Let $\mathbf{f}_i$ denote the force on ion $i$, and $\mathbf{F}=\lbrace{\mathbf{f}_1, \mathbf{f}_2, \ldots, \mathbf{f}_N \rbrace}$ is the 3N dimensional vector of all ionic force components.  Because of the large number of force components, we spend this section describing how we reduce the dimensionality of this data set to construct a handful of force-error measures.  

One of the simplest measures of force errors we can devise is the mean absolute force error  $\langle |\delta \mathbf{f}^{DF} | \rangle_S$.  Intuitively, this is the ensemble average magnitude of the force error vector $\delta \mathbf{f}_i = \mathbf{f}^{DF}_i - \mathbf{f}^{QMC}_i$. 

We can do significantly better than establishing which functional has better forces on average.  Indeed, knowing the force on each atom gives significant insight into the system's local structure.  Rigorously, one can do this through calculating the ``potential of mean force" $w(r)$, which is directly related to the pair correlation function by $g(r)=\exp(-\beta w(r))$.  Where the proper distribution is not known or hasn't been sampled adequately, one can often approximate this quantity fairly well and reproduce the major features of the pair correlation function, as is routinely done in force-matching with classical potentials.

For each density functional, we will define the following error measure relative to QMC.  Let a $i_{\mu}$ and $j_{\nu}$ denote two particles of species $\mu$ and $\nu$ respectively.  Denoting $\mathbf{r}_{i_\mu j_\nu} = \mathbf{r}_{i_\mu} - \mathbf{r}_{j_\nu}$ we define , 
\begin{equation}
\langle \delta f^{DF}_{\mu-\nu}(r) \rangle = \frac{\int d\mathbf{R} e^{-\beta E^{PBE}(\mathbf{R})} \delta(r-r_{i_\mu j_\nu}) \, \hat{\mathbf{r}}_{i_\mu j_\nu} \cdot \delta \mathbf{f}^{DF}_{i_\mu}}{\int d\mathbf{R} e^{-\beta E^{PBE} (\mathbf{R})} \delta(r-r_{i_\mu j_\nu})}
\end{equation}
Based on this definition, if $\langle \delta f^{DF}_{\mu-\nu}(r) \rangle$ is positive (negative), it overbinds (underbinds) species of type $\mu$ and $\nu$ at a distance $r$.  

Note that we use $E^{PBE}$ in this definition, since our configurations are sampled from QMD using the PBE functional.  However, if we could replace $E^{PBE}$ with $E^{QMC}$, then $\langle \delta f^{DF}_{\mu-\nu}(r) \rangle_{QMC}$ would be related to the density functional error in the potential of mean force $ \delta w^{DF}_{\mu\nu}(r)$ (relative to the QMC distribution) by
\begin{equation}
\langle \delta f^{DF}_{\mu-\nu}(r) \rangle_{QMC} = -\frac{\partial}{\partial{r}} \delta w^{DF}_{\mu\nu}(r)
\end{equation}

In any case, given that $E^{QMC}$ and $E^{PBE}$ produce qualitatively similar distributions of ionic configurations, $\langle \delta f^{DF}_{\mu-\nu}(r) \rangle$ will enable us to see over which regimes a density functional overbinds or underbinds, which would give strong indications as to how the $g(r)$ would change based on density functional.




\section{Results}

\label{sec:results}

\subsection{Global Energetics}\label{sec:global_energy}
Suppose we are interested in assessing to what extent the average error in the total energy changes as a function of helium concentration.  Depending on the error scaling, this can affect the both the Helmholtz and Gibbs free-energies of mixing, and thus the location of the H+He immiscibility transition.  

To measure this quantity, we define a measure of ``global energetics" as follows:  for a given $\rho$, we build an aggregated test set $S'(\rho)$ which consists of all test sets at all helium concentrations with a given electronic density $\rho$.  We then choose $c^{DF}(\rho)$ to be the median of $\lbrace \delta E^{DF} \rbrace_{S'(\rho)}$.  With this definition for $c^{DF}$, we define the ``global energetic error" of the test set $S(\rho,x_{He})$ by using Eq. \ref{error_measure}, which for shorthand we will refer to as $\langle |\widetilde{\delta E^{DF}}| \rangle_{g,S}$.  

\begin{figure}[h]
    \includegraphics[scale=0.5]{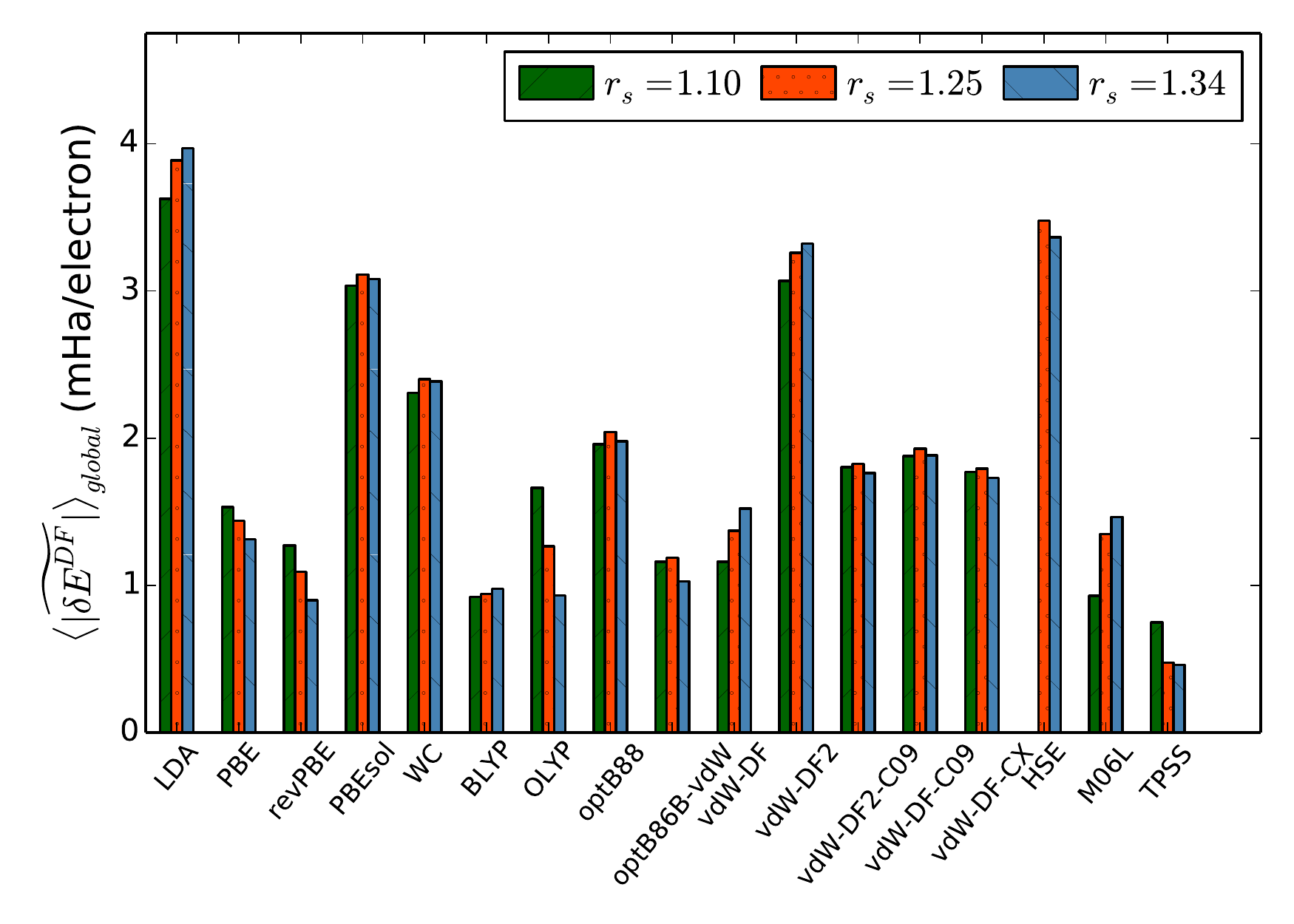}
   
     \caption{ $\langle |\widetilde{\delta E^{DF}}| \rangle_{global} $ averaged over all helium concentrations for all considered functionals.  The different bar colors/patterns denote the different densities.} 
     \label{fig:global_eerr_vs_func}
\end{figure}

In Fig. \ref{fig:global_eerr_vs_func}, we show the global energetic error for all functionals at three densities, averaged over all helium concentrations.  We see that the best performing functionals over all densities are the meta-GGA functionals TPSS and M06-L respectively, with global energetic errors that are under half that of PBE.  After these, the best performing functionals are the semi-local GGA's BLYP and revPBE, followed by optB86b-vdW and vdW-DF.  The worst performing functionals are LDA, HSE, PBEsol, and WC, with global energetic errors approximately twice that of PBE.  Though PBE has better than average performance in this regime, one can gain a decent fraction of a milihartree in accuracy by switching to a metaGGA or a properly tuned GGA.  

\begin{figure}[h]
    \includegraphics[scale=0.6]{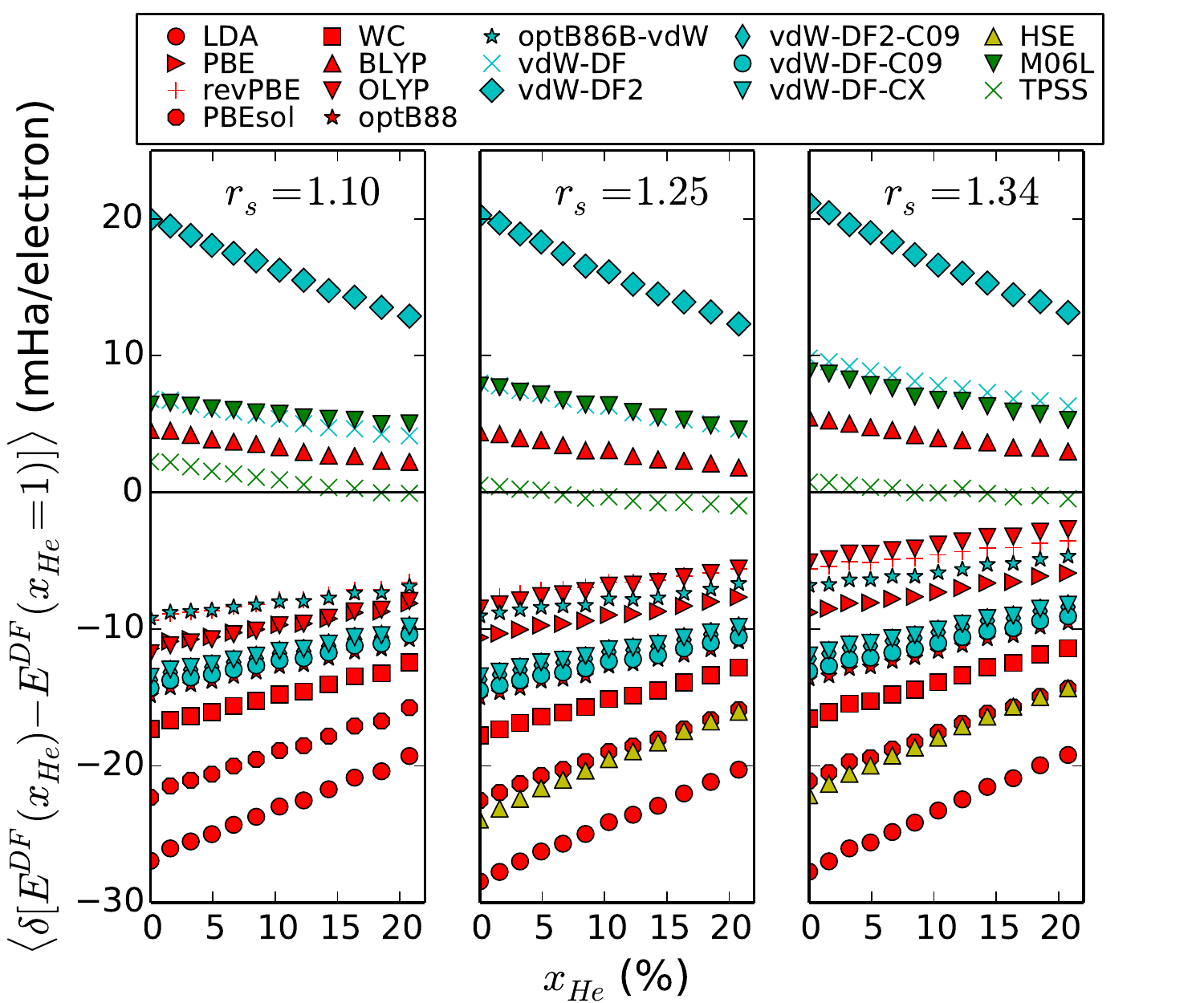}
   
     \caption{ $\langle \delta [ E(x_{He}) - E(x_{He}=1) ] \rangle $ vs. $x_{He}$ for all considered functionals at (left) $r_s=1.10$, (middle) $r_s=1.25$, and (right) $r_s=1.34$.  All energies are measured relative to the average energy of all pure helium configurations at the specified density. } 
     \label{fig:mean_eerr_vs_x_relHe}
\end{figure}

It is also useful to consider how accurately the energy difference between systems with different helium concentrations are captured with a density functional.  For specificity, we look at $\langle \delta [ E(x_{He},\rho) - E(x_{He}=1,\rho) ] \rangle $ in Fig. \ref{fig:mean_eerr_vs_x_relHe} for all considered functionals and helium concentrations. What we see is that relative to the pure helium configurations, BLYP, vdW-DF, and vdW-DF2 overestimate the energy difference between the mixed hydrogen/helium configurations, whereas all other functionals underestimate this difference.  Additionally, though all curves exhibit noticeable nonlinearity in the $x_{He}=0-20\%$ range, almost all curves have magnitudes that monotonically decrease to zero.  The exception is M06L, which appears as though it reaches a maximum energy error somewhere between $x_{He}=20-100\%$.


\subsection{Local Energetics}\label{sec:local_energy}

Even if the total energy error for an DFT-MD simulation at specific density and helium concentration averages to a small value based on the previous error measure, it is still possible for energy differences between similar configurations to have large errors.  To measure the spread of the error distribution, we use the following.  For each test set $S(\rho, x_{He})$ corresponding to a given density $\rho$ and helium concentration $x_{He}$, we set $c^{DF}$ to be the median of the set $\lbrace \delta E^{DF} \rbrace_{S(\rho, x_{He})}$.  Using Eq. \ref{error_measure} with this choice of test set and $c^{DF}$ gives our ``local energy" measure, which we denote as $\langle |\widetilde{\delta E^{DF}}| \rangle_{\ell,S}$.  Note that this is similar to the ``global energetic error" measure used previously, except now the test set $S$ and the reference set $S'$ are the same.  

\begin{figure}[h]
    \includegraphics[scale=0.5]{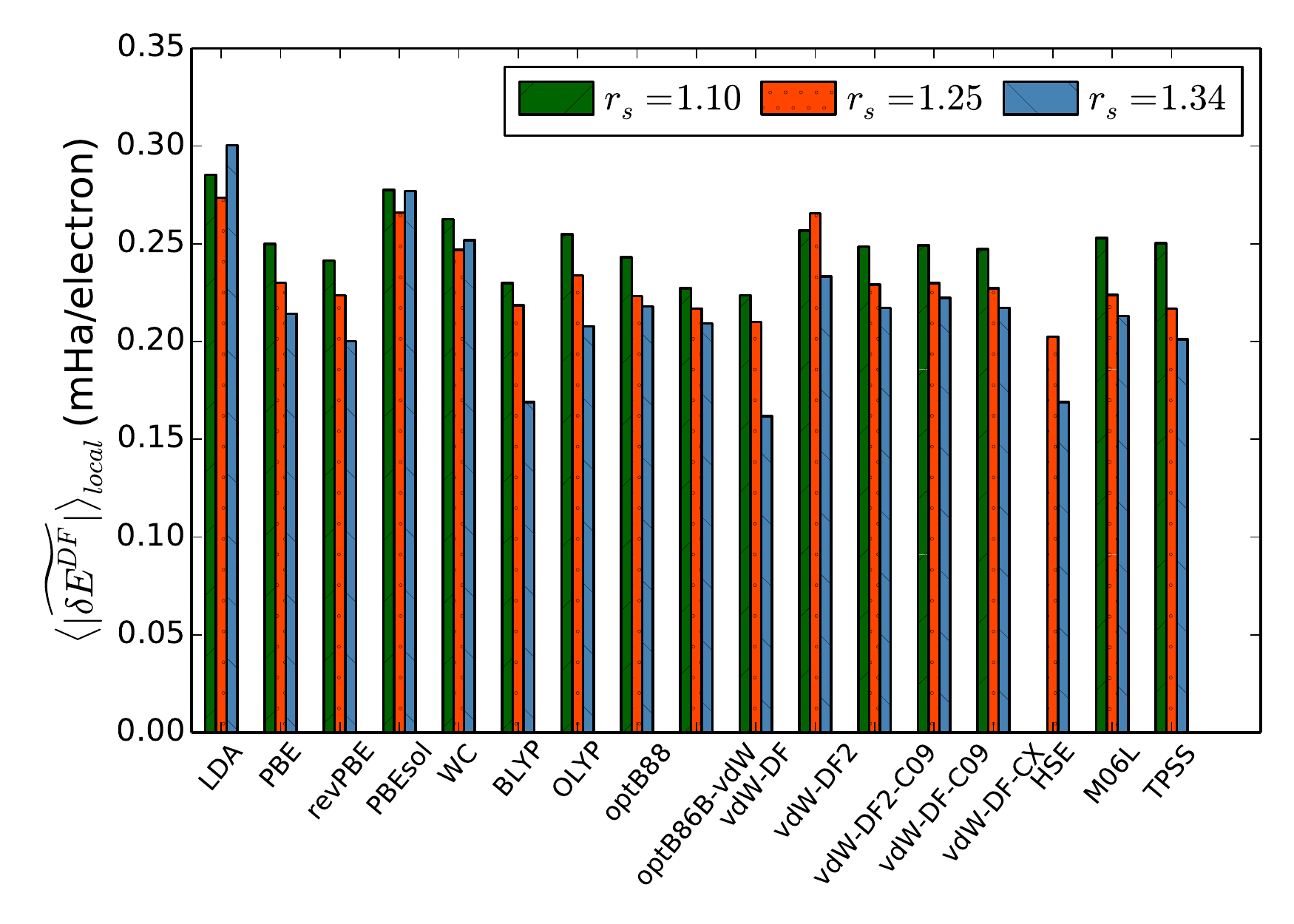}
    
     \caption{ $\langle |\widetilde{\delta E^{DF}}| \rangle_{local}$ averaged over all helium concentrations for all considered functionals.  The different bar colors/patterns denote the different densities.  } 
     \label{fig:local_energetics_vs_func}
\end{figure}


In Fig. \ref{fig:local_energetics_vs_func}, we show $\langle |\widetilde{\delta E^{DF}}| \rangle_{local}$ at three different densities for all functionals considered.  We note that the characteristic error scale is now ~0.25 mHa/electron, and that the differences between density functionals is significantly less pronounced than in the global energetic case.  So small are the differences, in fact, that little can be said for or against the vast majority of functionals by this metric, but the outliers are worth mentioning.  LDA, PBEsol, vdW-DF2, and M06L are the worst performers, whereas HSE is perhaps the best.  These trends are consistent with the global energetic trends, with the notable exception of M06L, which accurately captures global energetics, and HSE, which exhibits poor global energetic performance.  

\subsection{Pressures}\label{sec:pressure}
For all test sets and functionals, we computed $\langle \delta P^{DF}\rangle_S$ and $\langle \delta P^{DF} \rangle_S$.  As observed in pure hydrogen, we found the mean pressure errors to be systematically off, whereas $\langle |\delta P^{DF}| \rangle_S$ over all helium concentrations and densities is virtually indistinguishable from the statistical error bars.  $\langle \delta P^{DF}\rangle_S$ on the other hand can be quite sizable.  For the rest of this section, we will plot relative mean pressure errors instead of absolute errors because of the dramatic change in pressure as one changes the helium concentration.  For example, the pressure drops at $r_s=1.10$ from over $1TPa$ with pure hydrogen to around $300GPa$ for pure helium. 

\begin{figure}[h]
    \includegraphics[scale=0.5]{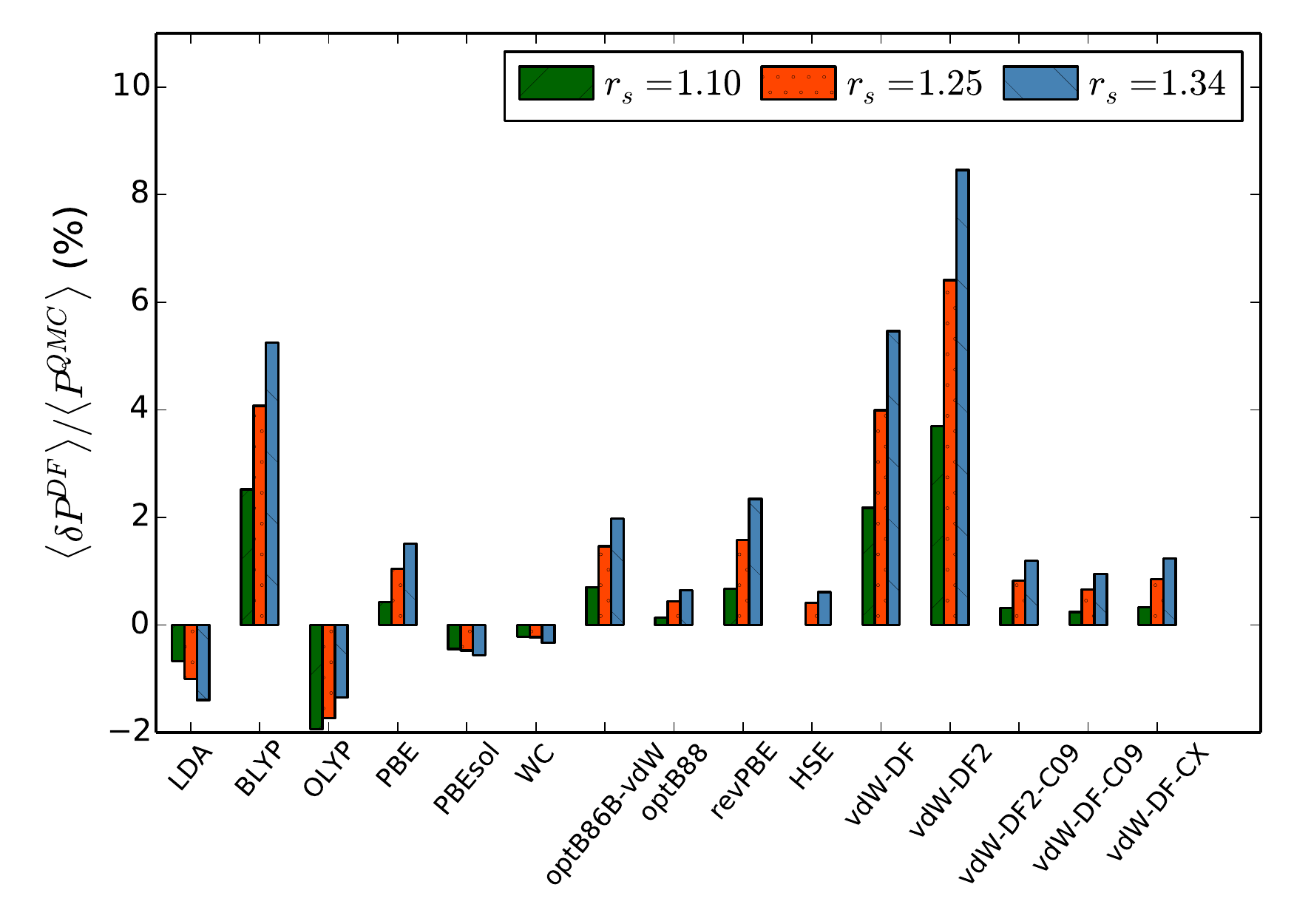}

     \caption{ $\langle \delta P^{DF} \rangle/\langle P^{QMC}\rangle$ in units of (\%) averaged over all helium concentrations for all considered functionals.  The different bar colors/patterns denote the different densities.  Not shown:  M06-L and TPSS.   } 
     \label{fig:avg_Perr_vs_func}
\end{figure}

In Fig. \ref{fig:avg_Perr_vs_func}, we plot $\langle \delta P^{DF}\rangle_S/\langle P^{QMC}\rangle$ (averaged over all helium concentrations) at three different densities for all functionals considered.  The trend observed is very much the same one observed in our previous benchmarking studies of pure hydrogen:  accurate energetics are compensated by poor pressure estimation.  For example, PBEsol, Wu-Cohen, LDA, though the worst performers energetically, provide the most accurate pressure estimates, missing the correct pressure estimate by less than 1\%.  The worst shown functionals happen to be vdW-DF and BLYP, which were known for their accurate energetics.  vdW-DF2 is a bit of an exception, in that it has poor energy and poor pressure estimation.  Note that TPSS and M06-L functionals are not plotted.  This is because relative to all other functionals, the pressure errors are quite significant.  Averaged over all three densities, M06-L and TPSS have average pressure errors of approximately -31\% and -17\% respectively, both exhibiting increasingly poor performance as the density is decreased.  

\begin{figure}[h]
    \includegraphics[scale=0.6]{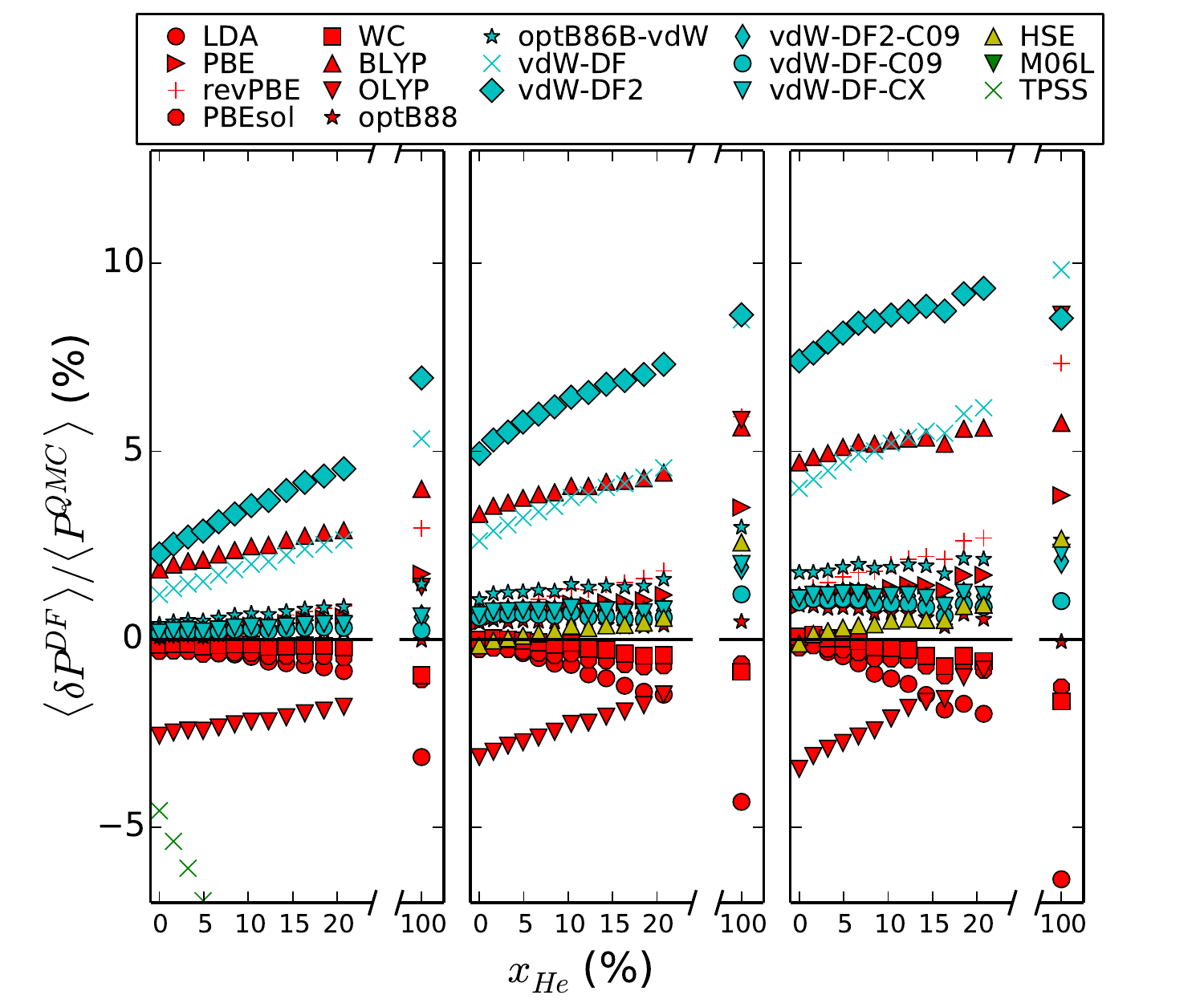}

     \caption{  $\langle\delta P^{DF} \rangle/\langle P^{QMC}\rangle$ in units of (\%) vs. $x_{He}$ for all considered functionals at (left) $r_s=1.10$, (middle) $r_s=1.25$, and (right) $r_s=1.34$.  The different colors/shapes are given in the legend and denote the density functional.  Note the broken axis, which shows the mean relative pressure error for the pure helium configurations. } 
     \label{fig:avg_Perr_vs_x}
\end{figure}

In Fig. \ref{fig:avg_Perr_vs_x}, we plot the mean pressure error versus helium concentration for all considered functionals.  Generically, though some functionals might underestimate or overestimate the pressure errors as the helium fraction is increased, the dependence of the pressure errors on  $x_{He}$ is very well behaved. The magnitudes for almost all functionals increases monotonically as $x_{He}$ increases, reaching its maximum error for pure He.  The exceptions are OLYP, where the pressure error has a positive slope and changes sign at some nonzero $x_{He}$, and WC, which reaches a maximum at some nonzero $x_{He}$ and then decreases towards a minimum at $x_{He}=1$.  Though we did not investigate helium concentrations higher than about 20\%, the smoothness of the pressure errors as a function of $x_{He}$ is reassuring, as it opens up the possibility of fitting these errors and correcting for them in post processing.  



\subsection{Enthalpies}\label{sec:enthalpies}
The mean enthalpy error $\langle \delta H^{DF} \rangle_S$ is given by $\langle \delta H^{DF} \rangle_S =\langle \delta E^{DF} \rangle_S+V \langle \delta P^{DF} \rangle_S $, which we can combine with the results from \ref{sec:local_energy} and \ref{sec:pressure} to study the errors in the predicted DFT enthalpy.

\begin{figure}[h]
    \includegraphics[scale=0.6]{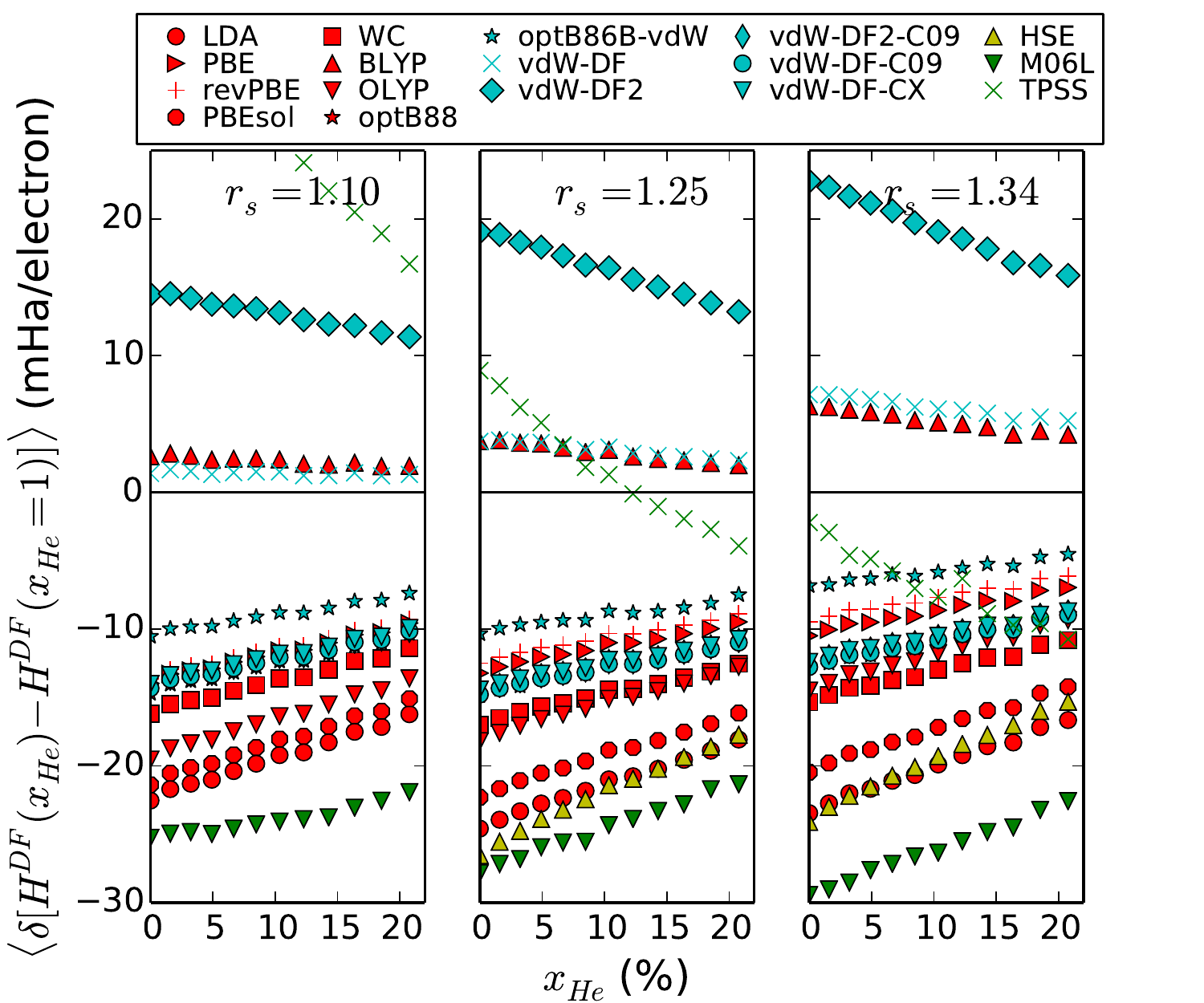}

     \caption{ $\langle \delta[ H^{DF}(x_{He})-H^{DF}(x_{He}=1)]\rangle $ vs $x_{He}$ for all considered functionals at (left) $r_s=1.10$, (middle) $r_s=1.25$, and (bottom) $r_s=1.34$.  All energies are measured relative to the average energy of all pure helium configurations at the specified density.  Note that the reference enthalpy for each density is taken to be the mean enthalpy of the pure helium configurations at that density.} 
     \label{fig:avg_entherr_vs_func}
\end{figure}

In Fig. \ref{fig:avg_entherr_vs_func}, we show the the average enthalpy error $\langle \delta H^{DF} \rangle_S$ versus helium concentration for $r_s=1.10,1.25,1.34$.  At fixed density $\rho$, $H^{DF}$ is measured relative to the enthalpy of pure helium at density $\rho$.  What we see is for the most part qualitatively similar to what we saw for the energy errors in Fig. \ref{fig:mean_eerr_vs_x_relHe}.  However, we see that the tested functionals possess varying degrees of error cancellation.  Some functionals definitely benefit from error cancellation:  specifically vdW-DF, LDA, and HSE.  This can reduce the absolute enthalpy by as much as 3-4 mHa/electron depending on the functional and density.  Others suffer from error \textit{addition}, such as PBE and most dramatically OLYP, which has an enthalpy error almost 10mHa/electron higher than its corresponding energy error.  Lastly, there are some functionals which exhibit neither error cancellation nor addition, namely BLYP and all the newer vdW functionals.  


\subsection{Forces}
\subsubsection{Total Force Errors}

\begin{figure}[h]
    \includegraphics[scale=0.5]{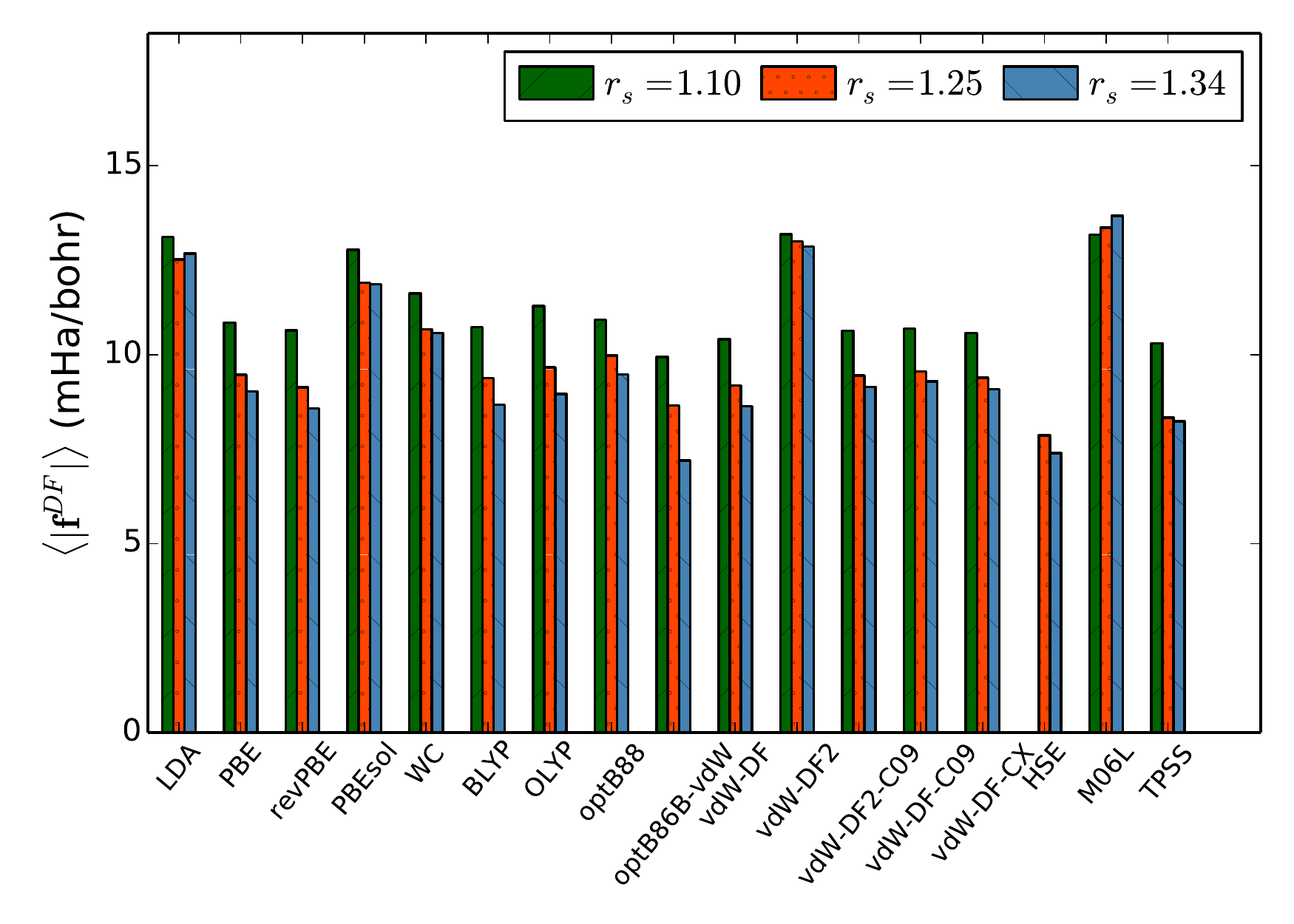}

     \caption{ $\langle | \delta \mathbf{f}^{DF}_H | \rangle $  aggregated over all helium concentrations for all considered functionals.  The different colors denote different densities. } 
     \label{fig:meanF_vs_func}
\end{figure}

The most natural question we can ask ourselves is which functional has the most accurate forces on average.  To this end, in Fig. \ref{fig:meanF_vs_func} we compute $\langle | \delta \mathbf{f}^{DF} | \rangle$ over all atoms and helium concentrations. Before interpreting Fig. \ref{fig:meanF_vs_func} however, we mention an important caveat.  Because of the large statistical noise relative to the individual force components, any mean absolute error measure is going to be saturated by the statistical noise.  Thus, the \textit{absolute magnitude} of the mean absolute force error should \textit{not} be interpreted as being indicative of the underlying functional performance.  In spite of this however, the fact that the statistical noise introduced in each force error measure is identical across all functionals, we are able to compare different functionals relatively and establish which ones are better and by how much.  


With the above caveats in mind, we note in Fig. \ref{fig:meanF_vs_func} that we observe the trend in mean absolute force errors is extremely similar to the trends we saw for local energetic errors.  Specifically, HSE is among the best, whereas vdW-DF2, LDA, and M06-L are among the worst. There are some slight differences however.  TPSS and optB86b-vdW seem to perform better by force error measures than the local energetics would suggest.  Additionally, though BLYP and vdW-DF still seem to outperform PBE, the difference is not nearly as apparent functionals as the local energetics would suggest.

\subsubsection{Local Force Errors}
In this section, we tabulate the average density functional force errors as a function of distance $\langle \delta f^{DF}_{\mu-\nu}(r) \rangle$, as described in Section \ref{sec:force_defs}.  We consider H-H, H-He, and He-He interatomic forces.  In principle, these force errors will not just depend on density functional, but also on density and helium concentration.  We will address these later two points first, as they will greatly simplify the analysis.  

\begin{figure}[h]
    \includegraphics[scale=0.6]{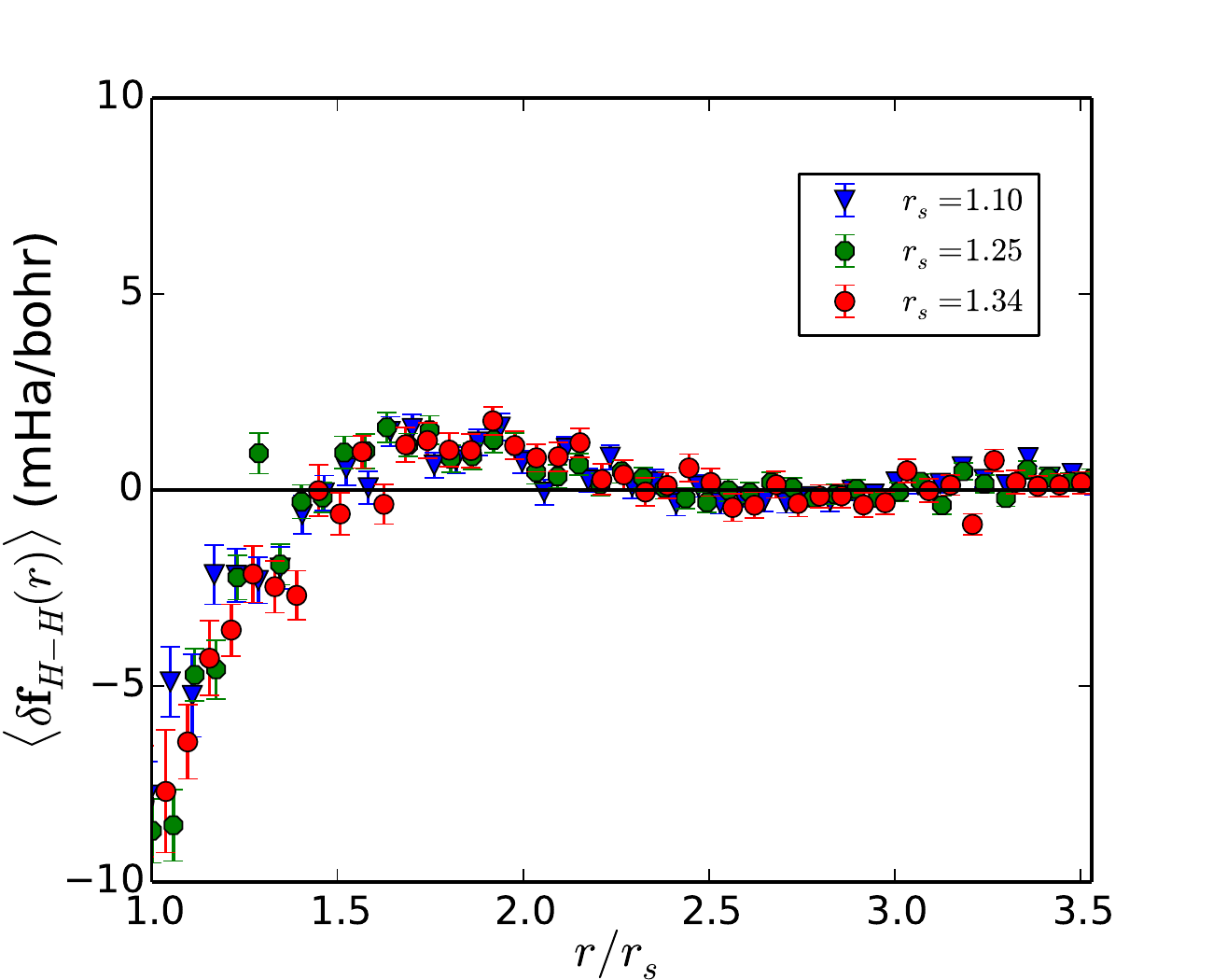}
    \includegraphics[scale=0.6]{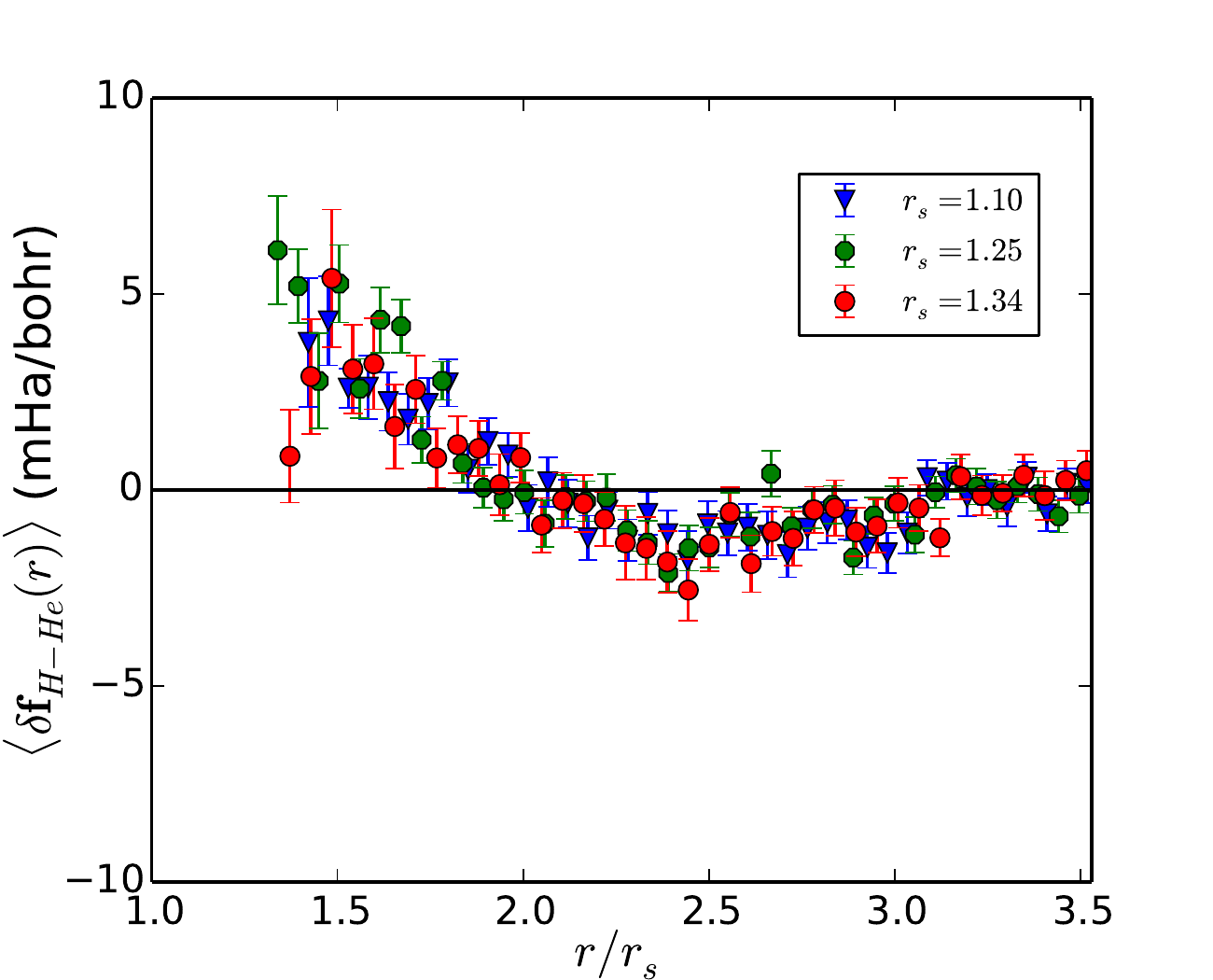}
    \includegraphics[scale=0.6]{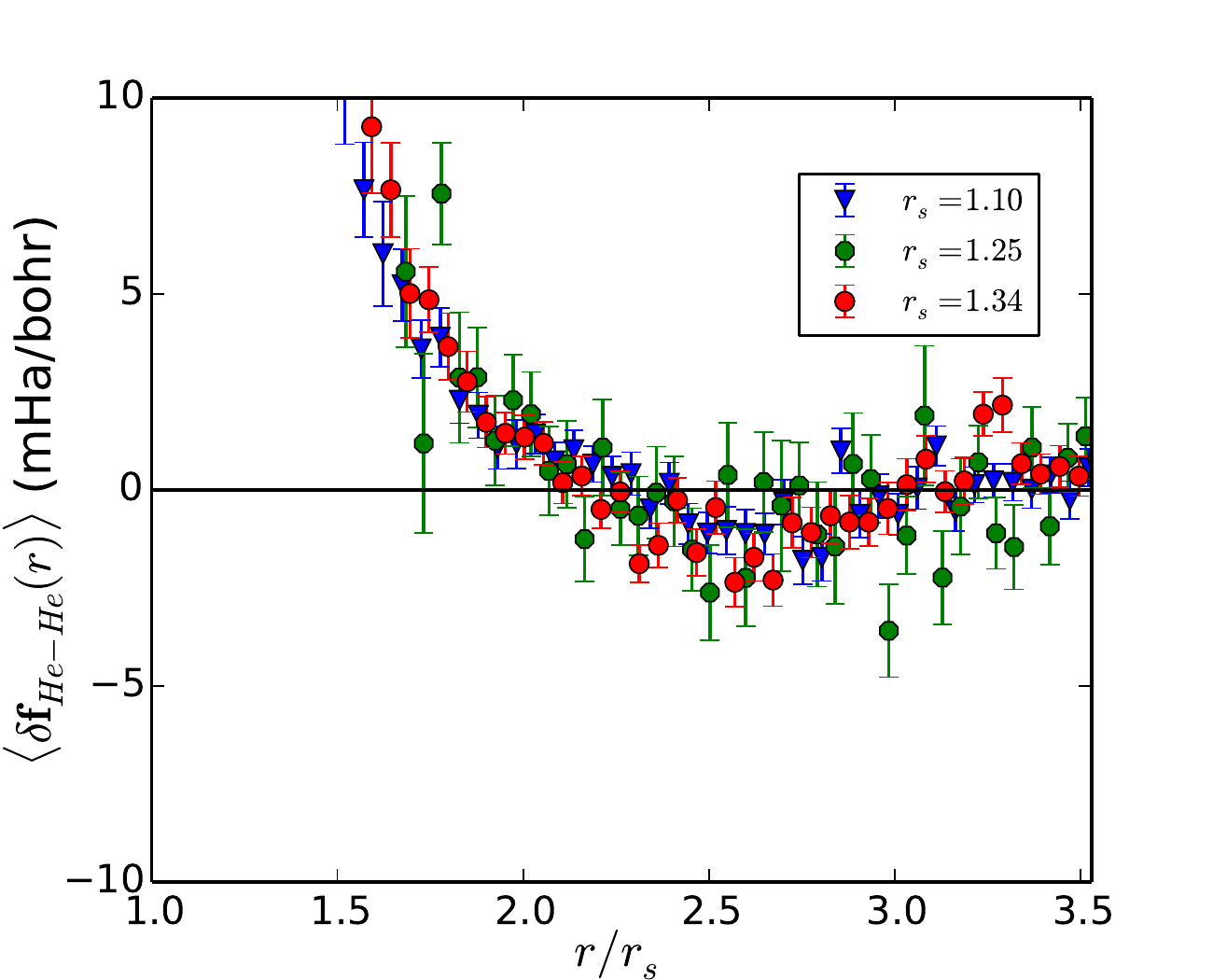}
     \caption{ $\langle \delta f^{PBE}_{\mu-\nu}(r) \rangle$ vs. $r/r_s$ as density is changed. The different marker colors/styles represent different densities.  (Top) $\langle \delta f^{PBE}_{H-H}(r) \rangle$ calculated at $x_{He}=20.7\%$, (middle) $\langle \delta f^{PBE}_{H-He}(r) \rangle$ calculated at $x_{He}=20.7\%$, (middle) $\langle \delta f^{PBE}_{He-He}(r) \rangle$ calculated at $x_{He}=100\%$} 
     \label{fig:fr_vs_density}
\end{figure}

The first question is how does the average force change as a function of density.  We show in Fig. \ref{fig:fr_vs_density} $\langle \delta f^{PBE}_{\mu-\nu}(r) \rangle$ versus $r/r_s$ calculated using the PBE functional.  We consider H-H (top), H-He (middle), and He-He (bottom) forces.  The first two were calculated at a helium fraction of 20.75\%, whereas the last was calculated in pure helium for statistical reasons.  For each plot, we overlay the plots of $\langle \delta f^{DF}_{\mu-\nu}(r) \rangle$ at the densities $r_s=1.10,1.25, 1.34$.  What is obvious is that with the exception of the sometimes large statistical fluctuations, no significant quantitative or qualitative differences exist in the mean force errors versus distance.  

\begin{figure}[h]
    \includegraphics[scale=0.6]{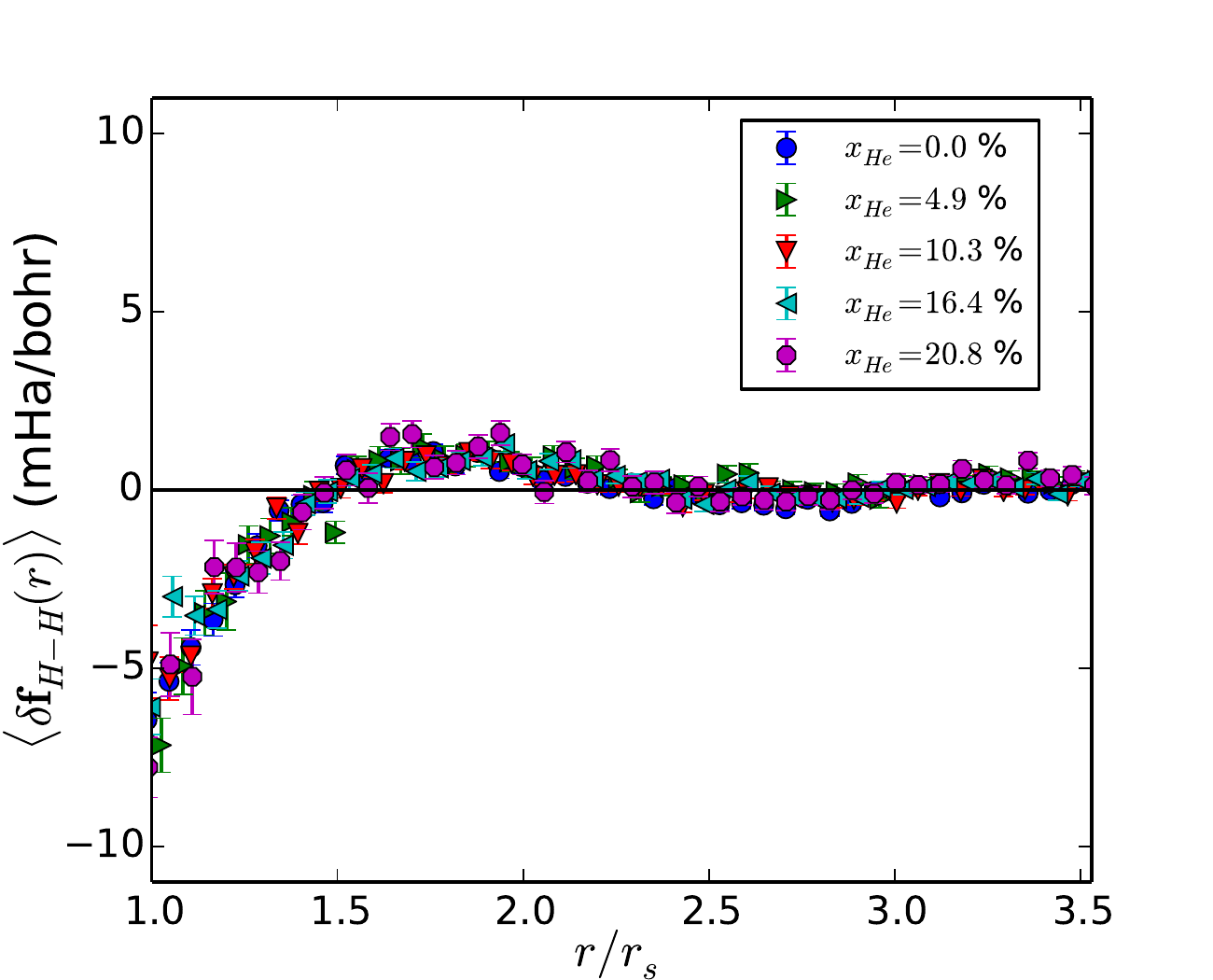}
    \includegraphics[scale=0.6]{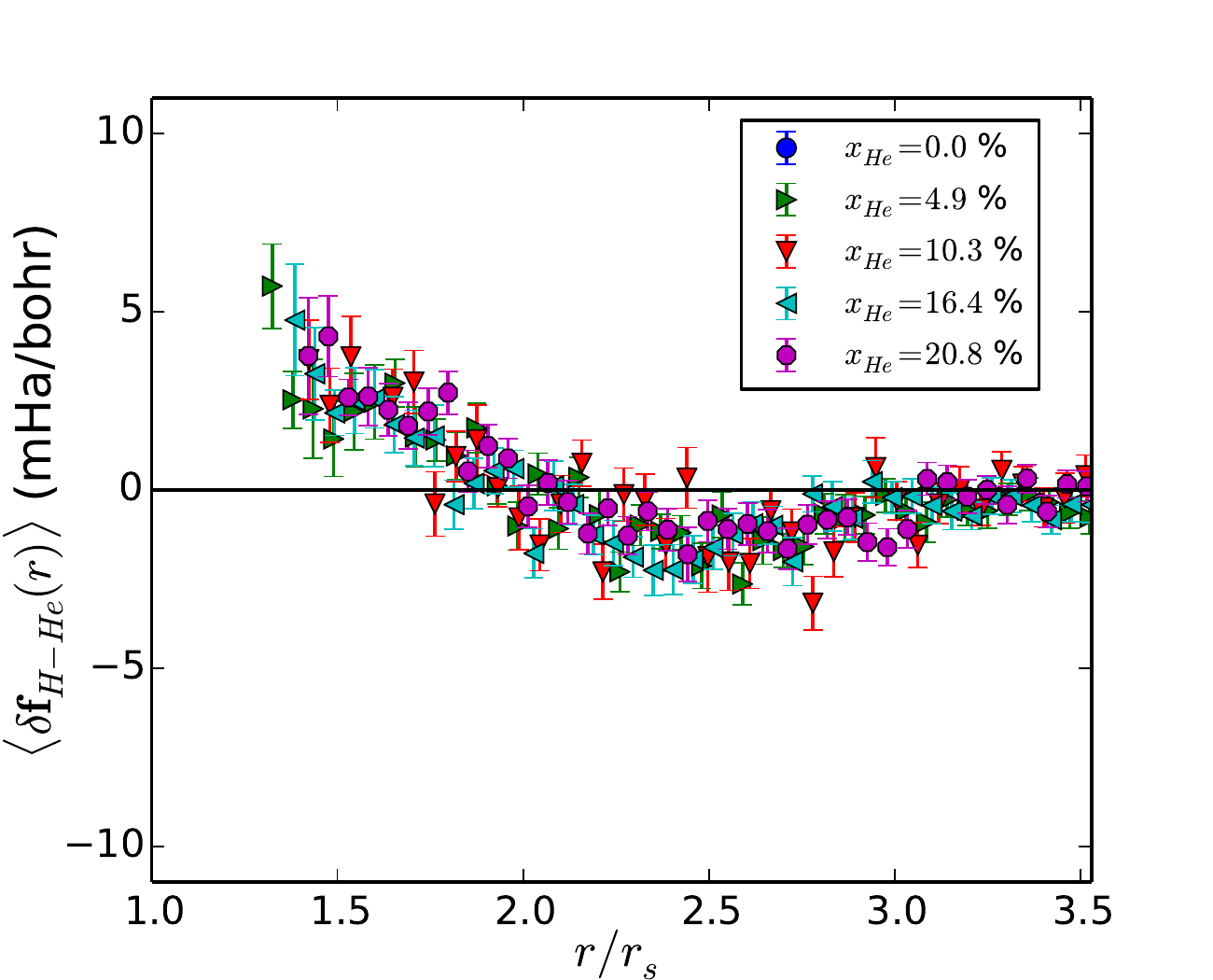}
     \includegraphics[scale=0.6]{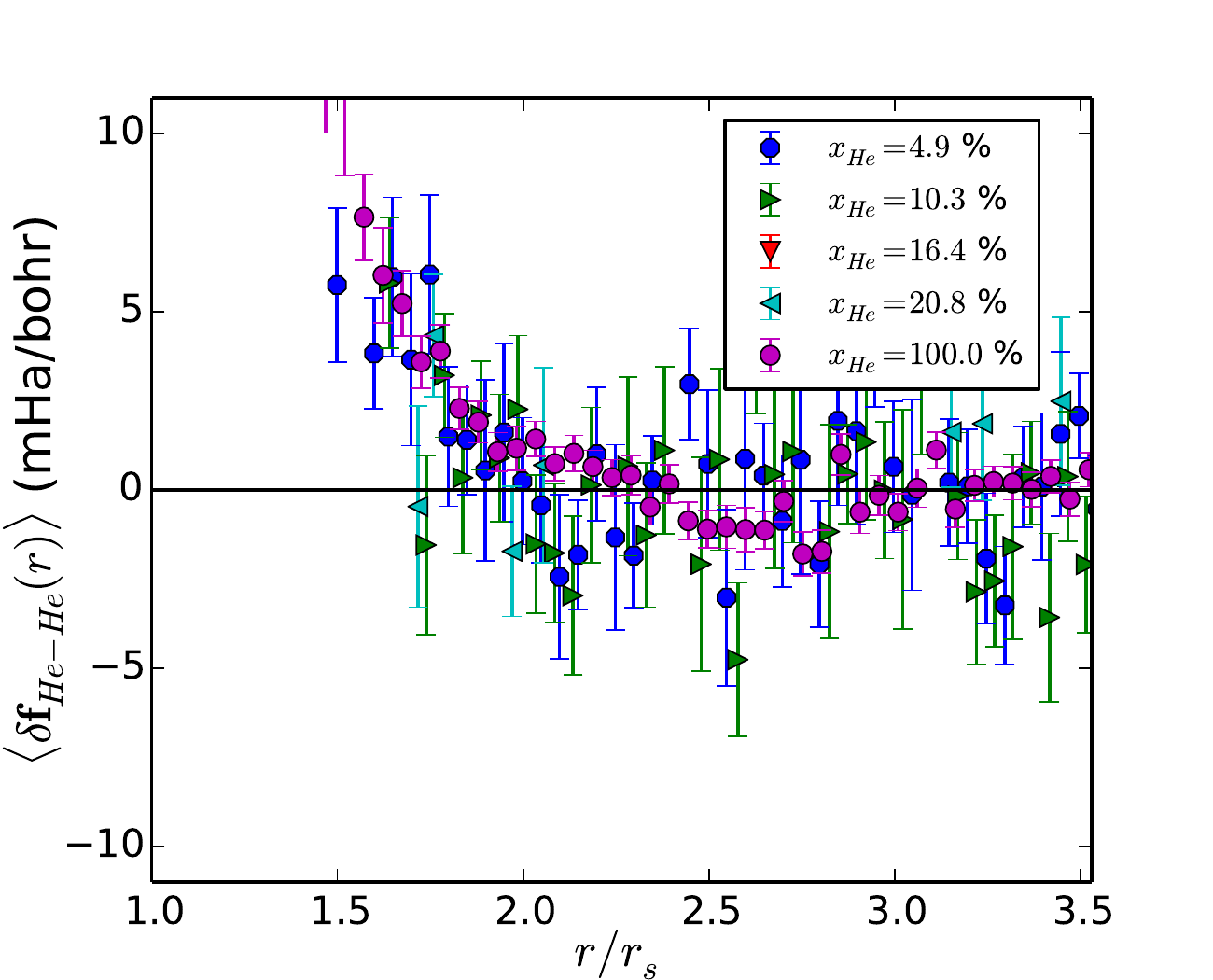}
     \caption{ $\langle \delta f^{PBE}_{\mu-\nu}(r) \rangle$ vs. $r/r_s$ as helium concentration is changed. The different marker colors/styles represent different helium concentrations. (Top) $\langle \delta f^{PBE}_{H-H}(r) \rangle$, (middle) $\langle \delta f^{PBE}_{H-He}(r) \rangle$, (middle) $\langle \delta f^{PBE}_{He-He}(r) \rangle$.  All configurations are at a density of $r_s=1.25$.} 
     \label{fig:fr_vs_hefrac}
\end{figure}

The second question is how does the average force change as a function of helium concentration.  In Fig. \ref{fig:fr_vs_hefrac}, we show the same general plots of $\langle \delta f^{PBE}_{\mu-\nu}(r) \rangle$ using the PBE functional as in Fig. \ref{fig:fr_vs_density}, but this time overlaying plots of different helium concentrations instead of different densities. All plots were calculated at a density of $r_s=1.25$.  Note that within error bars, $\langle \delta f^{PBE}_{\mu-\nu}(r) \rangle$ shows a remarkable insensitivity to helium concentration.

Given the insensitivity of the average force errors for PBE to both density and helium concentration, we plot $\langle \delta f^{DF}_{\mu-\nu}(r) \rangle$ for all considered functionals.  The helium fraction was chosen to be 1.6\% for the H-H (top), 20.75\% H-He (middle) plots, and 100\% for the He-He plot.  The density was chosen to be $r_s=1.25$.  Recalling from Section \ref{sec:force_defs} that $\langle \delta f^{DF}_{\mu-\nu}(r) \rangle > 0 $ implies overbinding relative to QMC.

\begin{figure}[h]
    \includegraphics[scale=0.6]{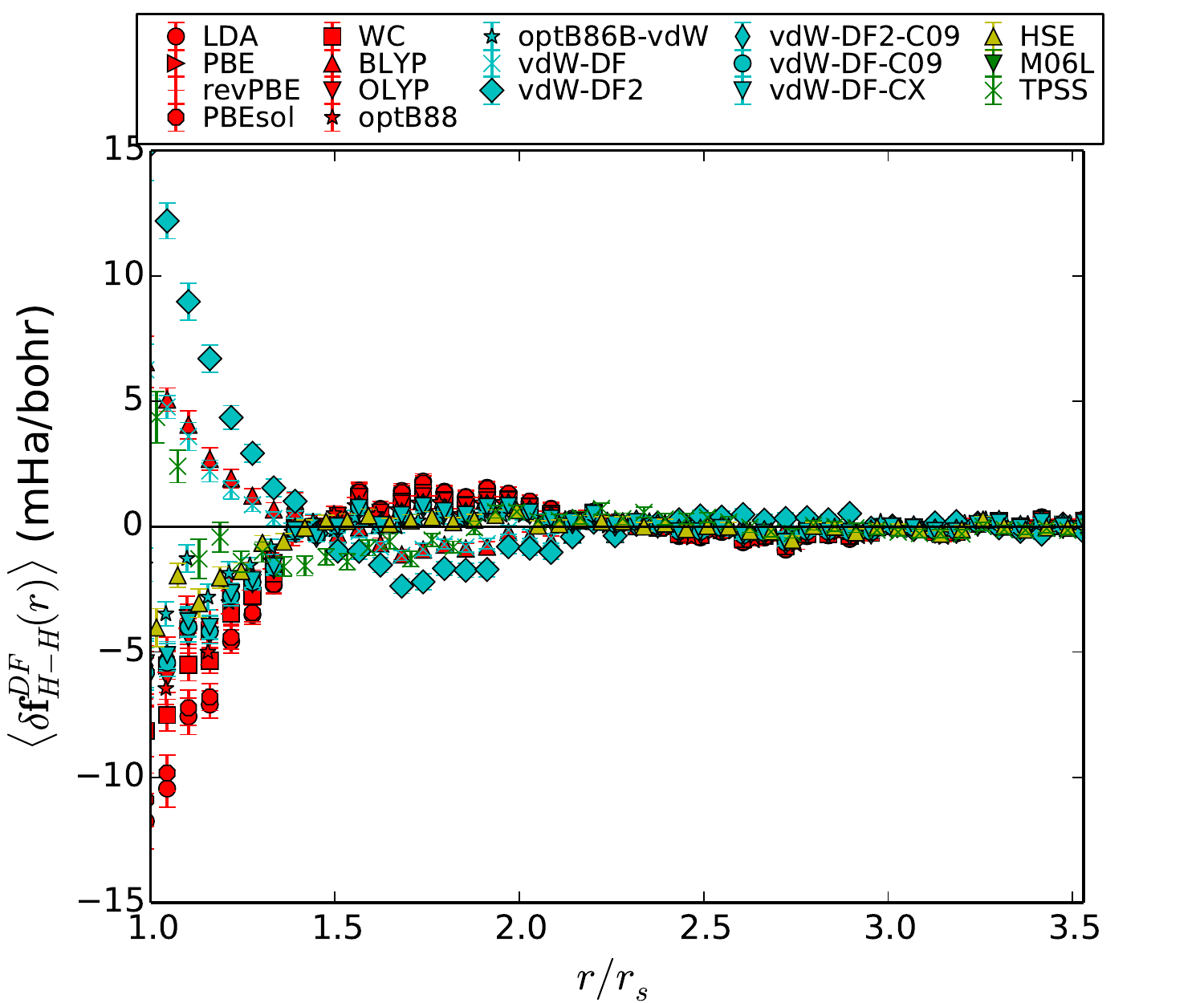}
    \includegraphics[scale=0.6]{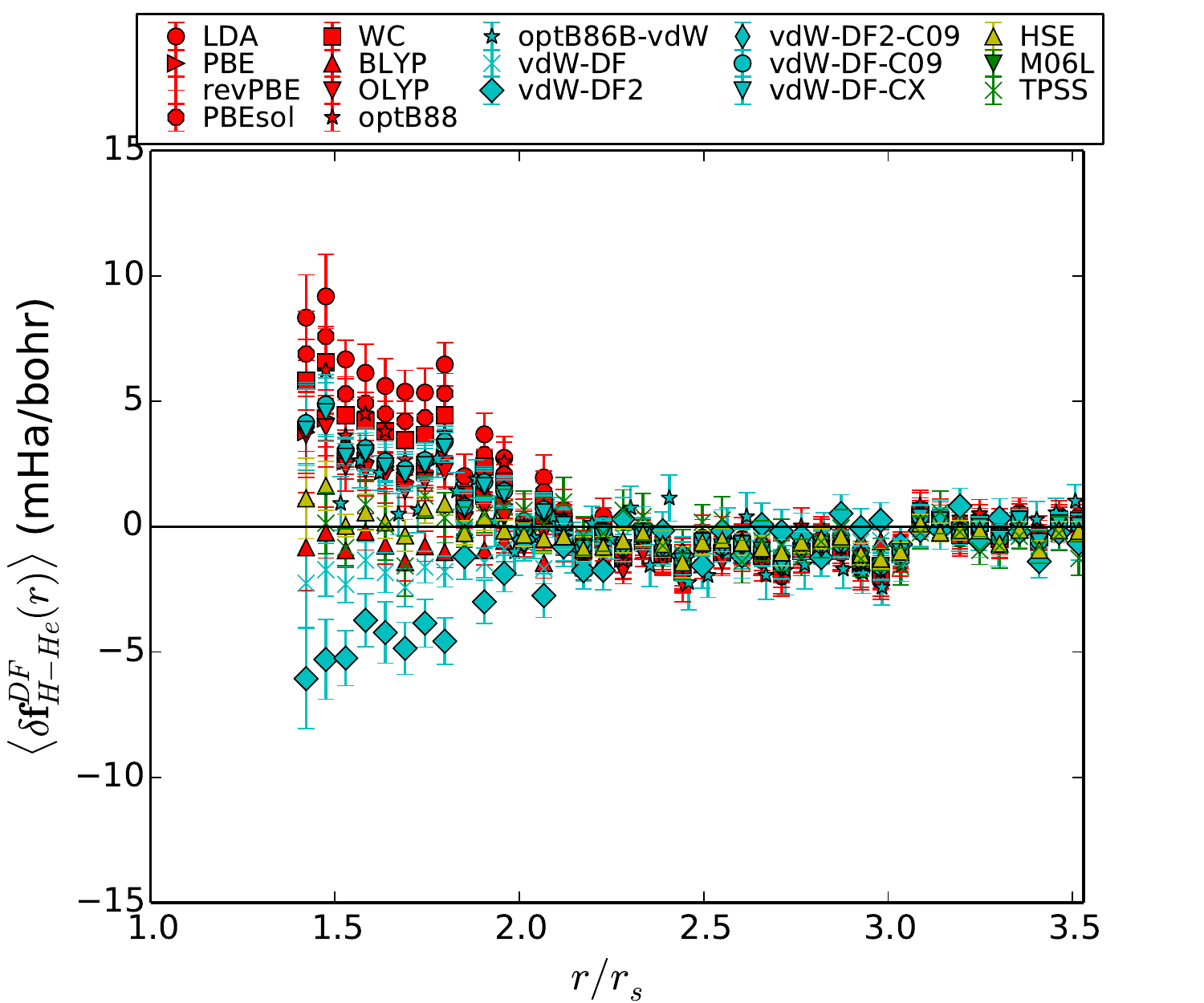}
    \includegraphics[scale=0.6]{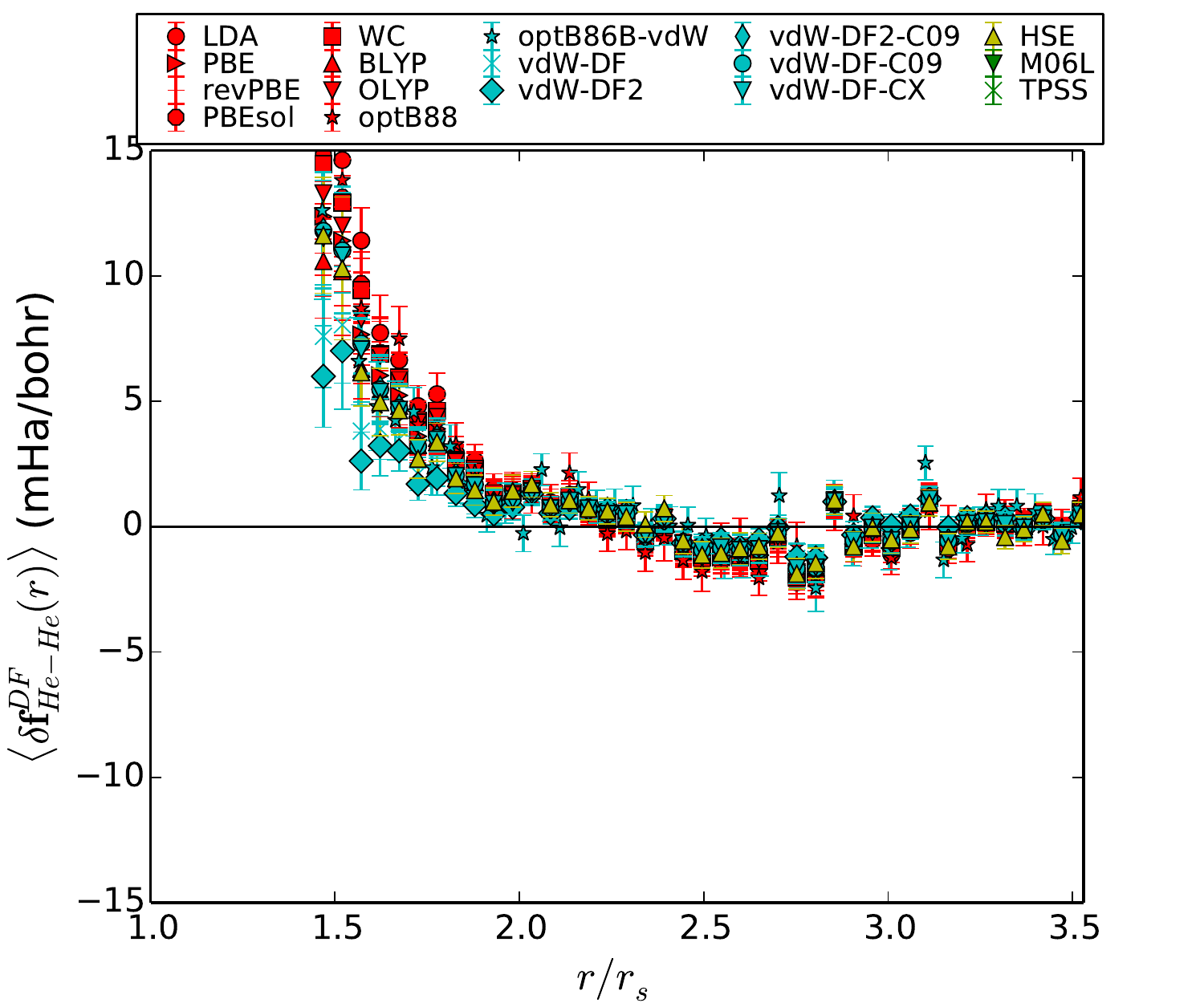}
     \caption{ $\langle \delta f^{DF}_{\mu-\nu}(r) \rangle$ vs. $r/r_s$ as the functional is changed. The different marker colors/styles represent different density functionals. (Top) $\langle \delta f^{PBE}_{H-H}(r) \rangle$ at $x_{He}=1.6\%$, (middle) $\langle \delta f^{PBE}_{H-He}(r) \rangle$ at $x_{He}=20.7\%$, (middle) $\langle \delta f^{PBE}_{He-He}(r) \rangle$ at $x_{He}=100\%$.  All configurations are at a density of $r_s=1.25$.} 
     \label{fig:mean_rad_F_vs_func}
\end{figure}

For the H-H forces at the top of Fig. \ref{fig:mean_rad_F_vs_func}, the BLYP, vdW-DF, and vdW-DF2 functionals all exhibit a strong propensity to overbind in the $1 < r/r_s < 1.5$, with vdW-DF2 overbinding the most.  TPSS overbinds the least in the region  $1 < r/r_s < 1.2$ but then underbinds slightly up to $r/r_s=2.2$.  All other functionals underbind in the region  $1 < r/r_s < 1.5$, with HSE underbinding the least and LDA the most.  Though its hard to tell with the noise, HSE has the lowest absolute error in the region $1 < r/r_s < 1.5$, followed by optB86b-vdW, vdW-DF and BLYP, and then by the combination revPBE, PBE, vdW-DF-CX, vdW-DF-C09, and vdW-DF2-C09.

For the H-H forces at the top of Fig. \ref{fig:mean_rad_F_vs_func}, there seem to be three distinct regions in space, whose boundaries are roughly where the vast majority of DF errors cross the $r$-axis.  I will refer to these as region I ($1 < r/r_s < 1.5$),  region II ($1.5 < r/r_s < 2.2$), and region III ($r/r_s>2.2$).  These should roughly correspond to the first, second, and third coordination shells.  Notice that with the exception of TPSS and M06-L, if a functional overbinds in region I, it will almost certainly underbind in region II, and overbind again in region III.  This is not entirely unexpected.  The ion-ion force depends only on electron densities, and so if two protons overbind because of an increased electronic charge between them, this decreases the electronic charge elsewhere, leading to underbinding in the charge depleted region.  

There are only a few functionals that overbind the H-H interaction in region I: BLYP, vdW-DF, and vdW-DF2, and TPSS (only for $r/r_s \approx 1$).  The rest underbind, though to varying degrees.  If we try to determine which functionals have the smallest error magnitudes in region I, we find that the trend is very similar to what we saw before in the mean absolute force and local energetic sections.  HSE, vdW-DF, BLYP, TPSS, and optB86b-vdW have the smallest errors in regions I, though further discrimination is difficult given the error bars.  In region II on the other hand, HSE and optB86b-vdW seem to have measureably smaller error magnitudes than vdW-DF, BLYP, and TPSS.  


For the H-He forces in the middle of Fig. \ref{fig:mean_rad_F_vs_func}, the differences between different functionals are more striking.  HSE and TPSS have the best average performance in the region $1.5 < r/r_s < 2.0$.  However, BLYP also performs exceptionally well, slightly underbinding hydrogen-helium pairs by less than 1mHa/bohr. The worst performing functionals are LDA, which overbinds the H-He interaction, and vdW-DF2, which underbinds.  

Lastly, we consider the He-He forces at the bottom of Fig. \ref{fig:mean_rad_F_vs_func}, the error bars are somewhat large, but there are some obvious trends still visible.  All functionals overbind the He atoms, although LDA overbinds the most. The functionals that underbind the least are either vdW-DF2, HSE, or TPSS, followed by vdW-DF and then maybe BLYP.

\section{Discussion}
\label{sec:discussion}
\begin{figure}
 \includegraphics[scale=0.6]{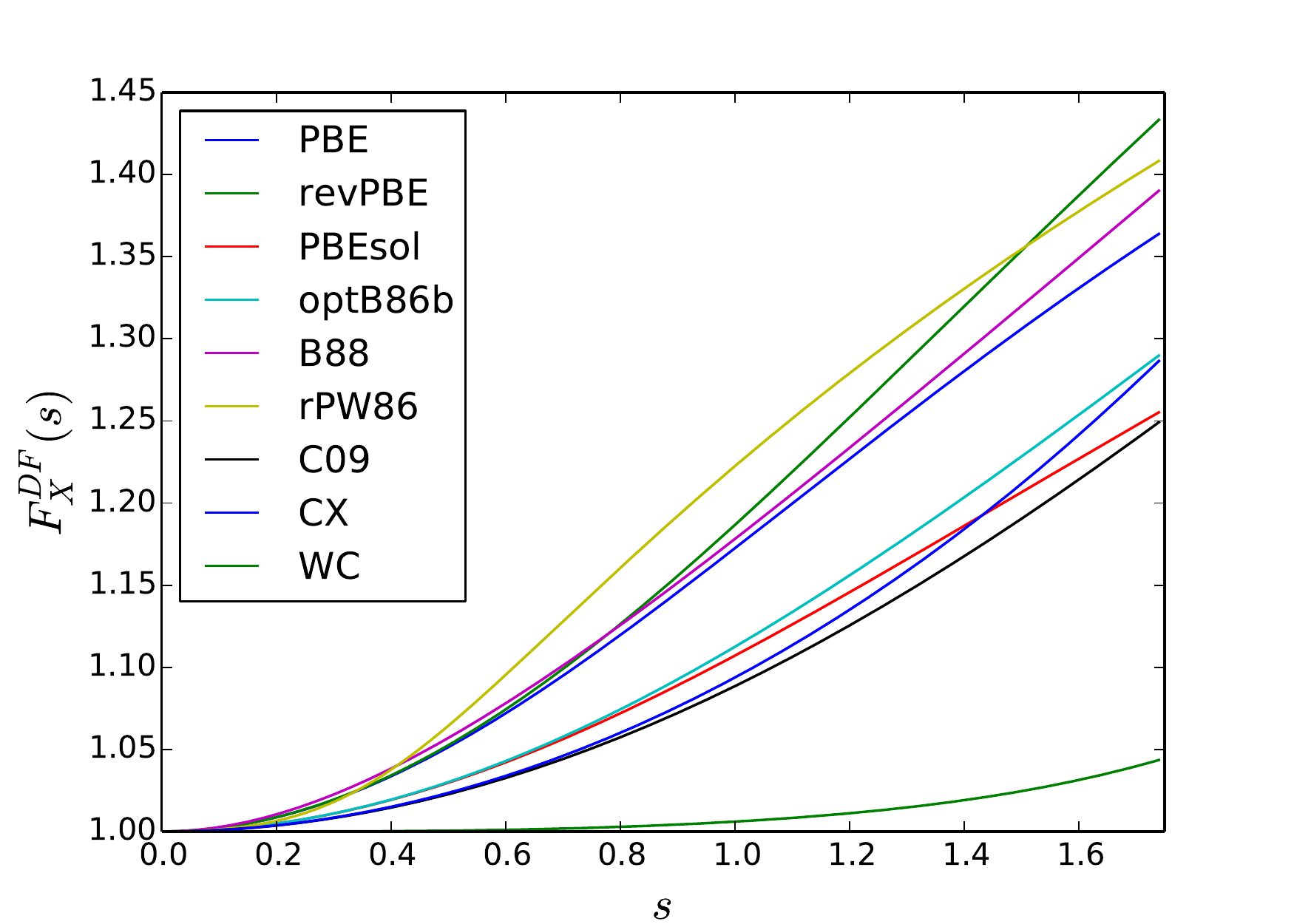}
 \caption{ Plot of enhancement factors $F_x(s)$ for all the best performing functionals. }
\label{fig:enhancement_factors}
\end{figure}

\subsection{Energeties}
In this section, we will try to reconcile the differences we observed between the local and global energetic trends.  We believe that the trends observed in the local energetic errors are mostly described by the impact that the density functional has on the charge density and forces, we will spend more time discussing that point in section \ref{sec:force_discussion}.  For now, we will try to grapple with how little bearing the local energetic errors had on the global energetic errors. 

The main point to realize is that the global energetic errors are going to be dominated by errors in the energy differences between configurations with different helium concentrations.  Calculating these types of energy differences accurately means one of two things: having small total energy errors, having error cancellation, or both.  Error cancellation is possible only if the H-H, H-He, and He-He interactions are described in roughly the same way.  But from our force discussion, we saw that many functionals will overbind one type of interaction while underbinding another.  An instructive comparison would be between vdW-DF and BLYP.  Both have very similar performances for forces and local energetics, with vdW-DF having a slight edge on both.  However, BLYP has noticeably smaller global energetic errors than vdW-DF.  Consider, for simplicity, the error in $E(x_{He}=0) - E(x_{He}=1)$.  Looking at the force errors, BLYP and vdW-DF overbind the H-H interaction in almost the exact same way.  However, vdW-DF underbinds the He-He interaction noticeably less than BLYP does.  The consequence is that the energy of the sampled pure He configurations (from PBE based QMD) are going to be \textit{higher} than vdW-DF will predict.  However, both vdW-DF and BLYP are overestimating the energy of a pure sampled H configuration by about the same amount, so the magnitude of $\delta [E(x_{He}=0) - E(x_{He}=1)]$ will be greater for vdW-DF than it will be for BLYP because of error cancellation.  This argument only goes so far however, as there are other contributions to the energy than just the electrostatic contribution.  HSE for instance, though probably describing the charge density very well and having small local energetic errors, has very poor global energetics.  


\subsection{Pressures}
As in our previous hydrogen benchmarking work, we observed that favorable energy errors were anti correlated with favorable pressure errors--the most dramatic cases being LDA and vdW-DF/BLYP.  Though this might seem paradoxical, the recognition of a tradeoff between accurate bulk moduli and lattice constants versus cohesive and atomization energies is well known.  In fact, Perdew et al. argue that this tradeoff is a necessary consequence of the limited form of the GGA functional \cite{Perdew2008}.  Basically, to reproduce accurate binding and atomization energies necessitates a $\mu$ that is higher than that associated with the gradient approximation expansion, which is necessary to recover the slowly-varying electron gas limit.  We find that almost all the best performing functionals in this regard are still the functionals that have been designed with these constraints in mind:  specifically, the LDA, Wu-Cohen, and PBEsol functionals.  

The hybrid functional HSE is the exception, instead using the reduced self-interaction error to achieve a better estimate of the pressure.  Not only did we observe reasonably accurate pressure estimation, HSE additionally had some of the smallest errors with local energetics and forces.  The global energetics errors were among the worst tested, but this might be due more to a more inequitable treatment of H and He and a lack of error cancellation, rather than a consequence of large absolute errors.  

\subsection{Enthalpies}
When constructing the equation of state for H+He mixtures, the most important thing we have to worry about is having accurate enthalpies.  Though the entropy term is also important, is far less sensitive to the choice of density functional than the energies and pressures.  That being said, one can cut the enthalpy errors by approximately 50-60\% relative to PBE (from ~11mHa/electron to ~4mHa/electron in pure hydrogen) by using either BLYP or vdW-DF.  Improving the enthalpy errors beyond this without using some sort of post-processing scheme might be somewhat difficult.  The 4mHa for vdW-DF and BLYP is in large part due to significant (though noticeably incomplete) error cancellation.  Given the inherent tradeoff between energy errors and pressure errors discussed previously, one should be extremely careful correcting each piece individually, especially if one can't fall back on a higher level of theory to verify.



\subsection{Forces}\label{sec:force_discussion}
From our analysis of local energetic errors and forces, we saw that there is a strong though not perfect correlation between small energetic errors and small force errors.  As mentioned before, having accurate electron-ion forces depends only on the ability to accurately reproduce the electronic charge density, whereas the local energetic error measures additionally fold in errors in the treatment of electron-electron correlation effects.

With this in mind, the superior performance of the HSE functional in minimizing the local energetic errors and force errors most likely stems from its ability to produce a reasonable charge-density.  This shouldn't be surprising, as the introduction of exact exchange favors charge localization through the reduction of self-interaction errors.  After HSE, TPSS seems to produce reasonable charge densities.  Among the GGA's and vdW corrected GGA's, the vdW-DF, BLYP, and optB86b-vdW functionals seem to produce reasonable charge densities, most likely because the underlying exchange functionals are skewed to energetically favor bonding and charge localization.

\subsection{Role of Exchange}

After benchmarking several GGA, meta-GGA, hybrid, and non-local vdW functionals, it is perhaps safe to say that most of the differentiation in performance between functionals we considered stems from the treatment of exchange, and not from the addition of sophisticated non-local correlation effects.  This conclusion follows from two pieces of evidence.  The first is that as far as global and local energetic errors are concerned, the two best performing functionals are vdW-DF, a non-local vdW functional, and BLYP, a GGA.  Beyond just having comparable performance, the total magnitude and scaling of local and global energy errors with helium concentration are very similar.  One would expect that if vdW type correlation were necessary for an accurate description of dense H+He, that there wouldn't be any GGA's performing nearly as well as the non-local vdW functionals.  

The other indication that the improved energetics is driven by the exchange piece comes from comparing the performance of the non-local van der Waals functionals.  vdW-DF, vdW-DF-C09, and vdW-DF-CX all use the same non local correlation functional, differing in their choice of exchange functional only.  The same is true of vdW-DF2 and vdW-DF2-C09.  We found that vdW-DF-C09 and vdW-DF2-C09 were virtually indistinguishable energetically, indicating the small role played by the difference in the van der Waals correlation piece.  However, there is a significant difference between vdW-DF-C09 and vdW-DF, or between vdW-DF2-C09 and vdW-DF2, each pair demonstrating either a propensity underbind or overbind respectively relative to QMC.  

It turns out that the best performing density functionals exhibit some common trends in their underlying exchange functionals.  The exchange functionals for GGA's are given by $E_x[\rho] = \int d\mathbf{r} \rho(\mathbf{r}) \epsilon^{hom}_{x}(\mathbf{r}) F_{x}(s(\mathbf{r}))$, where $ \epsilon^{hom}_{x}$ is the Slater-type exchange for the homogeneous electron gas, $F_{x}$ is the ``enhancement factor", and $s=|\nabla \rho |/[2 (3\pi^2)^{1/3} \rho^{4/3}]$ is the ``reduced density gradient".  Before getting into similarities in $F_x$ responsible for decent or poor energetic or pressure performance, we need to know which values of $s$ are relevant in our system.  After analyzing the PBE and BLYP charge densities for a single sample configuration from each density and helium concentration, we conclude that $s$ is bounded by $0<s<1.8$ for all configurations of interest ($s\leq 0.8$ for pure H configurations).  Unsurprisingly, the largest gradients occur in pure helium configurations at low density, whereas the smallest gradients occur in pure hydrogen at high density.  

Within the semilocal GGA functionals, we can explain better or worse energetic performance relative to PBE by looking at $F_x$.  We saw that the global energetic, local energetic, and force errors followed the progression of decreasing accuracy: BLYP, revPBE, PBE, and PBEsol.  Looking at the underlying enhancement factors (BLYP uses B88 exchange), we see the following trend:    $F^{B88}_x > F^{revPBE}_x > F^{PBE}_x > F^{PBEsol}_x$ for all ``s" in the relevant range for hydrogen.  $F^{revPBE}_x > F^{B88}_x$ from about $s=0.8$ onwards (they cross again at much larger s), but this doesn't affect the description of hydrogen.  This implies that the best performing functionals for energies and forces are working by lowering the energy contributions coming from the larger reduced gradients, which favors charge localization and bonding. Additionally, noting that the relative difference  $F_x^{B88}-F_x^{PBE} \approx 0.005 $ at $s=0.4$, and recalling how much the H-H forces and local energetics changed with respect to functional implies that the electronic structure around protons is \textit{very} sensitive to the treatment of exchange at these densities.  Looking at how similar the local energetic errors and He-He forces were for pure helium configurations for different functionals would indicate that the helium is not nearly as sensitive to the choice of exchange functional.

One can perform the same type of analysis with the vdW-DF type functionals.  vdW-DF, optB86b-vdW, and vdW-DF-CX use the revPBE, optB86b, CX exchange functionals respectively.  We previously saw that for energetic and force errors, the progression towards decreasing accuracy follows the sequence vdW-DF, optB86b-vdW, and vdW-DF-CX.  Looking at the underlying enhancement factors, we find that  $F^{revPBE}_x > F^{optB86b}_x > F^{CX}_x$.  vdW-DF and optB86b-vdW perform comparably, but vdW-DF overbinds relative to QMC whereas optB86-vdW underbinds.  We forgo a direct exhange functional comparison between the vdW functionals and the GGA's, primarily because of the ``exchange consistency" complication stemming from the use of a different ``outer" and ``inner" exchange correlation functional. 

Deeper relationships between the functional form of $F_x(s)$ and corresponding errors can be deduced from the previous discussion.  However, we leave these considerations to future publications, since our current focus is on hydrogen-helium thermodynamics and not on density functional development.

\section{Conclusions}
In this paper, we have used projector Quantum Monte Carlo to benchmark a collection of the most popular density functionals, ranging from GGA, to non-local dispersion corrected, to meta-GGA.  We were able to quantify the errors for most quantities that are relevant for constructing an equation of state:  specifically the pressures, local and global energy differences.  As a result of our analysis, we can conclude that significant reduction of enthalpy errors and a much better description of hydrogen helium interactions can be attained by using the TPSS metaGGA, the BLYP GGA, or the nonlocal vdW-DF functionals.   

Beyond just identifying the most accurate density functional and quantifying its errors, we have demonstrated the common features of the best performing functionals, specifically in the shape and limiting behavior of the enhancement factors for the exchange functionals.  The underlying exchange pieces for both vdW-DF, BLYP, and revPBE tend to emphasize bonding in the energetics, which is well known in the DFT literature.  The importance of this work is that it specifies quantitatively just how important this bonding character is for an accurate description of dense hydrogen helium mixtures.  Knowing this, and how much the various exchange correlation functionals overbind or underbind, should facilitate the optimization and deployment of new functionals for mapping out the H+He phase diagram.  


\begin{acknowledgments}
MAM was supported by the U.S. Department of Energy at the Lawrence Livermore National Laboratory under Contract DE-AC52-07NA27344.  MAM, RC and DMC were supported through the Predictive Theory and Modeling for Materials and Chemical Science program by the Basic Energy Science (BES), DOE. RC, and DMC were also supported by DOE DE-NA0001789.  MH and DMC acknowledge support from the Fondation Nanosiences de Grenoble.

Computer time was provided by the US DOE INCITE program, Lawrence Livermore National Laboratory through the 7th Institutional Unclassified Computing Grand Challenge program and PRACE project n 2011050781.

\end{acknowledgments}

\bibliography{benchmark} 

\end{document}